\documentclass[a4paper,11pt]{article}
\pdfoutput=1
\usepackage{jheppub}
\usepackage[T1]{fontenc}
\usepackage{verbatim}   
\usepackage{color}      
\usepackage{subfigure}  
\usepackage{multirow}
\usepackage{comment}
\usepackage{pdflscape}
\usepackage{rotating}
\usepackage{float}	
\newcommand{\lag}{{\cal L}}
\newcommand{\FDF}{\left(\varphi^\dagger\overleftrightarrow{D}_\mu\varphi\right)}
\newcommand{\FDFI}{\left(\varphi^\dagger\overleftrightarrow{D}^I_\mu\varphi\right)}

\title{Probing top quark neutral couplings in the Standard Model Effective Field Theory at NLO QCD}

\author[a]{Olga Bessidskaia Bylund,}
\author[b]{Fabio Maltoni,}
\author[b]{Ioannis Tsinikos,}
\author[b]{Eleni Vryonidou}
\author[c]{and Cen Zhang}
\affiliation[a]{Oskar Klein Centre and Department of Physics,\\ Stockholm University, SE-10691 Stockholm, Sweden} 
\affiliation[b]{Centre for Cosmology, Particle Physics and Phenomenology (CP3),\\
	Universit\'e catholique de Louvain, B-1348 Louvain-la-Neuve, Belgium} 
\affiliation[c]{Department of Physics, Brookhaven National Laboratory, Upton, NY 11973, USA}

\emailAdd{olga.bylund@cern.ch}
\emailAdd{fabio.maltoni@uclouvain.be}
\emailAdd{ioannis.tsinikos@uclouvain.be}
\emailAdd{eleni.vryonidou@uclouvain.be}
\emailAdd{cenzhang@bnl.gov}

\abstract{
 Top quark pair production in association with a $Z$-boson or a photon at the
 LHC directly probes neutral top-quark couplings.  We present predictions for
 these two processes in the Standard Model (SM) Effective Field Theory (EFT) at
 next-to-leading (NLO) order in QCD.  We include the full set of CP-even
 dimension-six operators that enter the top-quark interactions with the SM
 gauge bosons.  For comparison, we also present predictions in the SMEFT for
 top loop-induced $HZ$ production at the LHC and for $t\bar{t}$ production at
 the ILC at NLO in QCD.  Results for total cross sections and differential
 distributions are obtained and uncertainties coming from missing higher
 orders in the strong coupling and in the EFT expansions are discussed.   
 NLO results matched to the parton shower are available, allowing
 for event generation to be directly employed in an experimental analyses. 
 Our framework provides a solid basis for the interpretation of current and future measurements
 in the SMEFT, with improved accuracy and precision. }

\preprint{CP3-16-03, MCnet-16-03}

\begin{document}
\maketitle

\section{Introduction}
\label{sec:intro}

Top quark measurements are an important priority in Run II at the LHC. Results
from the Tevatron and the first run of the LHC at 7 and 8 TeV have yielded
precise measurements of the main top quark production channels,
i.e.~top--anti-top production and single top production.  At the LHC, the high
energy and luminosity open up new possibilities to access rarer production
processes, such as the associated production of top pairs with a vector boson.
These processes are particularly interesting, as they provide the first probe
of the neutral couplings of the top quark to the electroweak gauge bosons,
which were not accessible at the Tevatron due to their high production
thresholds.  Therefore these channels could give important information about
the top quark, which are complementary to top-pair and single-top production
measurements as well as the top decay measurements.  $t\bar t\gamma$ has been
measured at the Tevatron by CDF \cite{Aaltonen:2011sp}, and at the LHC
by CMS \cite{CMS:2014wma} and by ATLAS \cite{Aad:2015uwa}.  Results for
$t\bar{t}Z$ and $t\bar{t}W$ by CMS appear in \cite{Khachatryan:2014ewa,Khachatryan:2015sha} 
and by ATLAS in \cite{Aad:2015eua}.

Measurements of these processes allow us to search for deviations from the
Standard Model (SM) predictions.  While these deviations are often interpreted
in terms of anomalous top couplings, the SM Effective Field Theory (SMEFT)
provides a much more powerful framework
\cite{Weinberg:1978kz,Buchmuller:1985jz,Leung:1984ni}.  In this approach
possible deviations can be consistently and systematically described by the
effects of higher-dimensional operators of the SM fields.  By employing global
analyses~\cite{Durieux:2014xla,Buckley:2015nca,Buckley:2015lku}, experimental
results can be used to determine the size of the deviations due to each
effective operator.  The established deviations can then be consistently
evolved up to high scales, and matched to possible new physics scenarios.  In
the absence of convincing evidence for new resonance states, the EFT provides
the most model-independent approach to a global interpretation of measurements.

With Run-II of the LHC, more and more precise measurements in the top-quark
sector can be expected. In this respect, theoretical predictions matching the
foreseeable precision of the experimental determinations are required to
extract correct and useful information about deviations in the top-quark
sector.  For this reason, recently fully differential NLO QCD corrections to
top-quark processes within the top quark EFT have started to become available,
for example for the top-decay processes including the main decay channel and the
flavor-changing channels \cite{Zhang:2013xya,Zhang:2014rja}, and for
single-top production triggered by flavor-changing neutral interactions of the
top \cite{Degrande:2014tta}. More recently, the two main production channels
in the SM, top-quark pair production and single top production, have also become
available at dimension-six at NLO in QCD \cite{Franzosi:2015osa,Zhang:2016omx}.
QCD corrections are found to have nontrivial impact on SMEFT analyses
\cite{Zhang:2016omx}.

In this work we pursue this line of research further.  We provide NLO QCD
predictions for the $t\bar tZ$ and $t\bar t\gamma$ channels at the LHC and $t\bar t$
production at the ILC, including the full set of dimension-six operators that
parametrise the interactions between the top-quark and the SM gauge bosons. 
Note that results for $pp\to t\bar t\gamma$ at NLO appear here for the first time, while 
$pp\to t\bar tZ$ and $e^+e^-\to t\bar t$ have been
calculated at NLO in QCD in Refs.~\cite{Rontsch:2014cca,Rontsch:2015una} in the
anomalous coupling approach, albeit with the omission of the chromomagnetic dipole operator.  As we will see, this operator gives a very important contribution
to both the $t\bar tZ$ and $t\bar t\gamma$ processes. In addition, we also present
results for the top-loop induced $HZ$ production, which involves the same operators. 
An important feature of our approach is that NLO predictions matched to the 
parton shower (PS) are provided in an automatic way. Our
results are important not only because predictions are improved in accuracy and
in precision, but also because NLO results can be used directly in an
experimental simulation, allowing for a more dedicated investigation of all the
features of any potential deviations, with possibly optimised selections and improved
sensitivities to probe EFT signals.  Our approach is based on the
{\sc MadGraph5\_aMC@NLO} ({\sc MG5\_aMC}) \cite{Alwall:2014hca} framework, and
is part of the ongoing efforts of automating NLO EFT simulations for colliders~\cite{Zhang:2016snc}.    

The paper is organised as follows. In section \ref{sec:effoper} we present the
relevant dimension-six operators. In section \ref{sec:calcsetup} we present our
calculation setup.  Results for the $t\bar{t} Z$, $t\bar{t}\gamma$, $gg\to HZ$ processes at
the LHC and $t\bar{t}$ production at the ILC are given in sections
\ref{sec:resttV}-\ref{sec:ILC}, followed by a discussion about theoretical
uncertainties in section \ref{sec:unc}.  In section \ref{sec:disc} we discuss
the sensitivity of the various processes on the operators in light of the
corresponding LHC measurements. We draw our conclusions and discuss the outlook
in section \ref{sec:conc}.

\section{Effective operators}
\label{sec:effoper}

In an EFT approach, SM deviations are described by higher-dimensional operators.
Up to dimension-six, we consider the following operators
\cite{AguilarSaavedra:2008zc,Grzadkowski:2010es}:
\begin{flalign}
	&O_{\varphi Q}^{(3)} =i\frac{1}{2}y_t^2 \FDFI (\bar{Q}\gamma^\mu\tau^I
	Q) \label{eq:Ofq3} \\ &O_{\varphi Q}^{(1)} =i\frac{1}{2}y_t^2 \FDF
	(\bar{Q}\gamma^\mu Q) \\ &O_{\varphi t} =i\frac{1}{2}y_t^2 \FDF
	(\bar{t}\gamma^\mu t) \\
	&O_{tW}=y_tg_w(\bar{Q}\sigma^{\mu\nu}\tau^It)\tilde{\varphi}W_{\mu\nu}^I
	\\ &O_{tB}=y_tg_Y(\bar{Q}\sigma^{\mu\nu}t)\tilde{\varphi}B_{\mu\nu} \\
	&O_{tG}=y_tg_s(\bar{Q}\sigma^{\mu\nu}T^At)\tilde{\varphi}G_{\mu\nu}^A\,,
	\label{eq:Otf} 
\end{flalign} 
where $Q$ is the third generation left-handed quark doublet,  $\varphi$
is the Higgs field,  $g_W$, $g_Y$ and $g_s$ are the SM gauge coupling constants, 
$y_t$ is the top-Yukawa coupling, defined by $y_t=\sqrt{2}m_t/v$ where $v$
is the Higgs vacuum expectation value and $m_t$ is the pole mass (and so $y_t$
does not run).  At lowest order in perturbation expansion, the Lagrangian is modified by these operators as follows:
\begin{equation} \Delta \lag=\frac{C_{\phi Q}^{(3)}}{\Lambda^2}(O_{\phi
	Q}^{(3)}+h.c.)+ \frac{C_{\phi Q}^{(1)}}{\Lambda^2}(O_{\phi
	Q}^{(1)}+h.c.)+\dots \,,
\end{equation} 
i.e.~the Hermitian conjugate of each operator is added.  

The above operators form a complete set that parameterises
the top-quark couplings to the gluon and the electroweak gauge bosons of the SM,
which could contribute at $\mathcal{O}(\Lambda^{-2})$.
In this work we focus on their contributions to top production processes at colliders calculated at NLO in QCD.
The first three operators are tree-level generated current-current operators.
They modify the vector and axial-vector coupling of the
top quark to the electroweak gauge bosons.  The other three are dipole
operators, that are more likely to be loop induced.
$O_{tW}$ and $O_{tB}$ give rise to electroweak dipole moments, and $O_{tG}$ is
the chromomagnetic dipole operator, relevant for the interaction of the top
quark with gluons.  Up to order $\Lambda^{-2}$, the cross sections and
differential observables considered in this work do not receive CP-odd
contributions, so in the following we assume the coefficients of $O_{tW,tB,tG}$
to be real.  The three current operators are Hermitian so their coefficients are
always real.

A complete study of the processes considered here involve more operators at
dimension-six.  For example, four-fermion operators featuring top-quark pairs will
also contribute to these processes. They are the same set of seven operators
that contribute to top pair production as discussed in
\cite{Zhang:2010dr,Degrande:2010kt}.  Additional four-fermion operators could
enter and modify the $t\bar tZ$ vertex through loops. In this work we will not
consider this kind of operators, postponing this to future studies.
Operators involving the gauge bosons and light quarks could in principle
contribute to these processes, but as they receive stringent constraints from
precision observables, we consider their effect to these processes to be negligible
compared to the top operators. Another operator that contributes to the
$t\bar{t}Z/t\bar{t}\gamma$ processes is $O_G$ which would enter by modifying
the gluon self-interactions.  As this is not a top-quark operator,  we will not
consider it further here, assuming also that its contribution is sufficiently
suppressed due to constraints from the accurately measured $t\bar{t}$ and dijet
cross sections.

In our approach we also take into account an additional operator, $O_{\varphi
b}$ (identical to $O_{\varphi t}$ with $b$ replacing $t$), which does not
involve a top quark, but does contribute to, for example, NLO
$t\bar{t}Z$ production through a bottom loop or $b-$quarks in the initial state as well as 
$HZ$ production in gluon fusion through the bottom loops. We include it in this
study mainly as an option to cancel the $ggZ$ chiral anomaly induced by
modifications to the $ttZ$ interaction.

Various constraints can be placed on the Wilson coefficients of the top quark
operators of Eqs.~(\ref{eq:Ofq3}-\ref{eq:Otf}) both from direct measurements and
from electroweak precision measurements. For $\Lambda=1$ TeV, at $95\%$ confidence
level, $C_{tG}$ is constrained from top pair production to be within the range
[-0.77,0.4] in Ref.~\cite{Aguilar-Saavedra:2014iga}, and in
Ref.~\cite{Franzosi:2015osa} [-0.56,0.41] at LO and [-0.42,0.30] at NLO.  $C_{tW}$
is constrained from $W$ helicity fractions in top-decay measurements and single
top production, to be in the interval [-0.15,1.9]
\cite{Tonero:2014jea}. The $Z\to b \bar{b}$ decay constrains the sum of
$C_{\phi Q}^{(3)}+ C_{\phi Q}^{(1)}$ to be [-0.026, 0.059] \cite{Zhang:2012cd}.
The other three operator coefficients, $C_{\phi Q}^{(3)}-C_{\phi Q}^{(1)}$,
$C_{\phi t}$ and $C_{tB}$ receive indirect constraints from precision
electroweak data, which lead to the following limits
\cite{Greiner:2011tt,Zhang:2012cd}:
\begin{eqnarray*} 
	&&C_{\phi Q}^{(3)}- C_{\phi Q}^{(1)}:\ [-3.4, 7.5]\\ 
	&&C_{\phi t}:\ [-2.5, 7]\\ 
	&&C_{tB}:\ [-16, 43] \,.
\end{eqnarray*} 
Note that indirect bounds should be interpreted carefully.  The presented
bounds here are marginalised over the $S$ and $T$ parameters, with all other
operator coefficients assumed to vanish. We note here that comparable limits
have been set on these operators by the recent collider based global analyses
of \cite{Buckley:2015nca,Buckley:2015lku}.  Furthermore, RG-induced limits are
also can be found in \cite{deBlas:2015aea}.

Finally, let us stress that even though we work in the context of the SMEFT,
the NLO calculations presented in this work can be directly used in analyses
employing an anomalous couplings parametrisation, under the condition that
$C_{tG} = 0$ is assumed at all scales. In this case, operators do not mix, and
they only contribute via anomalous couplings in $ttV$, $bbV$ and $tbW$
vertices, and our NLO results can be translated into the anomalous coupling
approach. The relations between the anomalous couplings and the effective
operator coefficients are given in appendix A. 

\section{Calculation setup}
\label{sec:calcsetup}

Our computation is performed within the {\sc MG5\_aMC} framework
\cite{Alwall:2014hca}, where all the elements entering the NLO computations
are available automatically starting from the SMEFT Lagrangian \cite{Alloul:2013bka, Degrande:2011ua,Degrande:2014vpa, Hirschi:2011pa,Frederix:2009yq,Hirschi:2015iia}.  NLO results can be matched to parton shower programs, such as  PYTHIA8 \cite{Sjostrand:2014zea} and HERWIG++ \cite{Bahr:2008pv},  through the {\sc MC@NLO} \cite{Frixione:2002ik} formalism. 

Special care needs to be taken for the UV and R2 counterterms, which are required for the virtual corrections.  The R2 terms are
obtained automatically through the {\sc NLOCT} package \cite{Degrande:2014vpa}, and have been
checked against analytical calculations.  The UV counterterms depend on the
renormalisation scheme.  For the SM part, we use $\overline{MS}$ with
five-flavor running of $\alpha_s$ with the top-quark subtracted at zero
momentum transfer.  The bottom quark mass is neglected throughout.  Masses and wave-functions are renormalised on shell.
The operator $O_{tG}$ gives additional contributions to the top-quark and gluon
fields, as well as $\alpha_s$ renormalisation \cite{Franzosi:2015osa}.
The operator coefficients are subtracted with the $\overline{MS}$ scheme.
They are renormalised by
\begin{equation}
	C_{i}^0\to Z_{ij}C_{j} = 
	\left(1+\frac{1}{2}\Gamma(1+\varepsilon)(4\pi)^\varepsilon
	\frac{1}{\varepsilon_{UV}}\gamma \right)_{ij}C_j \,,
	\label{eq:zc}
\end{equation}
where the anomalous dimension matrix $\gamma$ has non-zero components for
the dipole operators $O_{tG}$, $O_{tW}$, and $O_{tB}$.  The anomalous dimensions
for these three operators are \cite{Zhang:2014rja}
\begin{equation}
	\gamma=\frac{2\alpha_s}{\pi}
	\left(
	\begin{array}{ccc}
		\frac{1}{6} &0 &0\\
		\frac{1}{3} & \frac{1}{3} &0\\
		\frac{5}{9} &0 & \frac{1}{3}\\ 
	\end{array}
	\right)\,.
	\label{eq:adm}
\end{equation}
The other operators do not have  an anomalous dimension at order
$\mathcal{O}(\alpha_s)$ due to current conservation.  Results in this work
are presented in terms of operators defined at the renormalisation scale,
which we take as $m_t$ for $pp\to t\bar tV$ and $e^+e^-\to t\bar t$, and
$m_H$ for $pp\to HZ$.  If the operator coefficients are known at the new physics scale $\Lambda$,
the above anomalous dimension matrix can be used to evolve them down to 
the renormalisation scale, to resum the large $\log\Lambda/m_t$ terms.
Hence results presented in this work are free of such large log terms. In general, we
find that NLO results cannot be approximated using the renormalisation group equations of the operators.

Operators that modify the $ttZ$ axial coupling may induce a chiral anomaly in the $ggZ$
three point function, which has an effect in $t\bar tZ$ and $gg\to HZ$
production.  The cancellation of the anomaly depends on the details of the
underlying model. To cancel this anomaly within the EFT framework, one option is
to include the operator $O_{\phi b}$, which modifies the $bbZ$ coupling, and
require
\begin{equation}
	C_{\phi b}=2C_{\phi Q}^{(1)}-C_{\phi t}
\end{equation}
so that the change in $ttZ$ and $bbZ$ vertices cancel each other in the $ggZ$
function.  In this work, we keep this anomaly in the calculation, and take the
point of view of \cite{Preskill:1990fr}, i.e.~the chiral anomaly in an
effective theory is allowed, provided the corresponding gauge boson is massive.
We have checked that, in either case, the numerical effect is negligible.  Note that 
the $SU(3)_C$ gauge is not affected, and related Ward Identities have been
verified.

As a cross-check of our implementation we have compared our (LO) results with those presented in
Ref.~\cite{Rontsch:2015una}, and have found agreement.

\section{Results for $t\bar{t}Z$, $t\bar{t}\gamma$ and $t\bar{t}\mu^+ \mu^-$}
\label{sec:resttV}

\subsection{Inclusive $t\bar{t}Z$, $t\bar{t}\gamma$ and $t\bar{t}\mu^+ \mu^-$
results}
In this section, we consider the inclusive $t\bar{t}Z$, $t\bar{t}\gamma$
and $t \bar t l^+ l^-$cross sections including the dimension-six operators.
The $t \bar t l^+ l^-$ cross section includes the contribution of off-shell photons
and the interference of $t \bar t Z$ and $t \bar t \gamma^*$. In fact, this is the process
that is experimentally accessible at the LHC, though the difference between
$t \bar t l^+ l^-$ and $t \bar t Z$ with leptonic $Z$ decay is small for a lepton pair invariant
mass close to the $Z$ boson mass.

We work up to $\mathcal{O}(\Lambda^{-2})$, generating Feynman diagrams with at most one
effective vertex. The cross section can then be expressed in the form:
\begin{equation} 
	\sigma=\sigma_{SM}+\sum_i
	\frac{C_i}{(\Lambda/1\textrm{TeV})^2}\sigma_i^{(1)}+\sum_{i\le j}
	\frac{C_i C_j}{(\Lambda/1\textrm{TeV})^4}\sigma_{ij}^{(2)}\,,
\end{equation} 
with the sum running over all operators in Eqs.~(\ref{eq:Ofq3}-\ref{eq:Otf}).
Here $\sigma_i^{(1)}$ is the cross section of the interference of diagrams with
one EFT vertex with diagrams from the SM. The cross section $\sigma_{ij}^{(2)}$,
corresponds to the interference of two diagrams with one EFT vertex each or the squares of the amplitudes with one
effective vertex for $i=j$.

Our implementation allows the extraction of the $\mathcal{O}(\Lambda^{-2})$
contribution $\sigma_{i}^{(1)}$ as well as the $\mathcal{O}(\Lambda^{-4})$
contribution $\sigma_{ij}^{(2)}$. While the latter is formally higher-order
with respect to the $\mathcal{O}(\Lambda^{-2})$ accuracy of our computation in
the SMEFT, it is important for several reasons. First, as this term is of higher-order one
can decide to include it without changing the accuracy of the
prediction of the central value.  Arguments in favour of this approach in the
SMEFT have been put forward, see
e.g.~\cite{Domenech:2012ai,Biekoetter:2014jwa}.  Finally, the
$\mathcal{O}(\Lambda^{-4})$ terms are useful to associate an uncertainty to
missing higher-orders in the EFT expansion.  For these reasons, we quote results
for $\sigma_{ii}^{(2)}$ (i.e.~the squared contribution from $\mathcal{O}_i$),
to either improve the central value predictions or
to (partly) assess the size of the theoretical uncertainties associated to the
contribution of $\mathcal{O}(\Lambda^{-4})$ and higher terms.

In this context, we point out that the relative size of  $\sigma_{ii}^{(2)}$ with respect to $\sigma_i^{(1)}$ cannot be
used to infer the breaking down of the EFT expansion which even in the case where $\sigma_{ii}^{(2)}\gg\sigma_i^{(1)}$ could
still be valid.  One reason is that $\sigma_i^{(1)}$ is an interference term and various
cancellations could occur accidentally.  We will see this is indeed the case
for several operators in $t\bar{t}V$ production.  On the other hand, the EFT expansion
in $E^2/\Lambda^2$ could still be well-behaved, or at least can be controlled by applying
kinematic cuts on the total energy $E$ of the process. In this respect, as we were mentioning above, a legitimate and motivated
way to proceed is to always include both interference and squared contributions, and
separately estimate the theoretical error due to missing dimension-eight
operators.  Another interesting possibility is in the presence of ``strong interactions", i.e.  when
$C_i^2 \frac{E^4}{\Lambda^4}>C_i\frac{E^2}{\Lambda^2}>1>\frac{E^2}{\Lambda^2}$.
In this case the squared contribution dominates over the interference one,
without invalidating the $E^2/\Lambda^2$ expansion, which is parametrically independent of the size of the coefficients.  
In a phenomenological analysis and in a global fit, all such cases should be always kept in mind and carefully analysed on the basis of  the resulting bounds on the $C_i$'s. 
As the main goal of this paper is to present a framework to perform calculations in the SMEFT at NLO accuracy and study the results
for the neutral top interactions, we do not discuss any further the issue related to the size of the coefficients and the validity conditions of the EFT itself. On the other hand, we stress that our implementation/framework can provide the elements necessary to make a detailed study. For example,  we present the full results at $\mathcal{O}(\Lambda^{-2})$, characterised by $\sigma_i^{(1)}$, together with $\sigma_{ii}^{(2)}$ as an estimation of uncertainties due to neglecting all $\sigma_{ij}^{(2)}$ terms. Note that if necessary, any $\sigma_{ij}^{(2)}$ term can be also computed.
 
In practice, to extract the values of $\sigma_i^{(1)}$, we set one of the $C_i$
coefficients to $\pm 1$ and all the others to zero. Using the two values and
the SM cross-section, we can extract $\sigma_i^{(1)}$, as well as
$\sigma_{ii}^{(2)}$, the contribution of the $\mathcal{O}(\Lambda^{-2})$
amplitudes squared. In order to improve the statistical significance of the
interference for the operators where the interference is small, we find the
value of $C_i$ which maximises it compared to the total cross-section and use
that value for the runs instead of $C_i=\pm 1$.
 
The results are obtained using the 5-flavour scheme, with the MSTW2008
\cite{Martin:2009iq} parton distribution functions. 
The input parameters are:
\begin{flalign}
	&m_t=173.3\ \mathrm{GeV}\,, \qquad
	m_Z=91.1876\ \mathrm{GeV}\,, \\
	&\alpha_{EW}^{-1}=127.9\,, \qquad
	G_F=1.16637\times10^{-5} \mathrm{GeV}^{-2}\,.
	\label{eq:input}
\end{flalign}
The renormalisation and
factorisation scales are fixed to $\mu_R=\mu_F=\mu=m_t$.  For a detailed
discussion of scale choices for the $t\bar{t}V$ processes see
\cite{Maltoni:2015ena}. Scale variations are obtained by independently setting
$\mu_R$ and $\mu_F$ to $\mu/2$, $\mu$ and $2\mu$, obtaining nine $(\mu_R,\mu_F)$
combinations. For the $t\bar{t}Z$ process no cuts are applied on the final
state particles and no $Z$ or top decays are considered, while for
$t\bar{t}\gamma$, $p_T(\gamma)>20$ GeV is required.  We employ the photon
isolation criterium of Ref.~\cite{Frixione:1998jh} with a radius of 0.4.
Finally for the $t\bar{t}\mu^+\mu^-$ process a cut of 10~GeV is set on the
minimum invariant mass of the lepton pair. 

\begin{table}[h]
\renewcommand{\arraystretch}{1.5}
\scriptsize
\begin{center}
\begin{tabular}{c c c c c}
\hline
 SM & [fb] & $t \bar t Z$ & $t \bar t \gamma$ & $t \bar t \mu^+ \mu^-$ \\
 \hline
8TeV & $\sigma_{SM,LO}$ & $ 207.0^{ +41.4 \% }_{ -26.8 \% }~^{ +2.4 \% }_{ -2.5 \% }$ & $ 604.0^{ +38.8 \% }_{ -25.6 \% }~^{ +2.1 \% }_{ -2.2 \% } $ & $ 8.779^{ +40.9 \% }_{ -26.6 \% }~^{ +2.4 \% }_{ -2.4 \% }$ \\
 & $\sigma_{SM,NLO}$ & $ 226.5^{ +6.7 \% }_{ -11.2 \% }~^{ +2.8 \% }_{ -3.2 \% }$ & $ 777^{ +13.4 \% }_{ -13.7 \% }~^{ +2.1 \% }_{ -2.4 \% }$ & $ 9.827^{ +7.7 \% }_{ -11.5 \% }~^{ +2.6 \% }_{ -2.9 \% }$  \\
  & K-factor & 1.09 & 1.29 & 1.12 \\
 \hline
13TeV & $\sigma_{SM,LO}$ & $ 761.8^{ +37.8 \% }_{ -25.2 \% }~^{ +2.1 \% }_{ -2.2\% }$ & $ 1998.0^{ +35.5 \% }_{ -24.2 \% }~^{ +1.8 \% }_{ -2.0 \% } $ & $ 31.67^{ +37.4 \% }_{ -25.1 \% }~^{ +2.1 \% }_{ -2.2 \% }$ \\
 & $\sigma_{SM,NLO}$ & $ 879^{ +8.0 \% }_{ -10.9 \% }~^{ +2.0 \% }_{ -2.5 \% } $ &  $ 2719^{ +14.2 \% }_{ -13.5 \% }~^{ +1.6 \% }_{ -1.9 \% } $  &  $ 37.51^{ +9.1 \% }_{ -11.3 \% }~^{ +2.0 \% }_{ -2.4 \% } $ \\
   & K-factor & 1.15 & 1.36 & 1.18 \\
\hline
\end{tabular}
\end{center}
 \caption{\label{tab:SM_8_13} SM cross sections (in fb) for $t \bar t Z$, $t \bar t \gamma$, $t\bar{t}\mu^+\mu^-$ production at the LHC at $\sqrt{s} =  8$~TeV and $\sqrt{s} =  13$~TeV.  The first percentage corresponds to scale variations and the second to PDF uncertainties. }  
\end{table} 
 
The SM predictions for the processes considered here are summarised as a
reference in Table \ref{tab:SM_8_13}, where  uncertainties from scale
variation, PDF uncertainties, and the K-factors are shown for the LHC at 8 and
13 TeV. The scale uncertainties are significantly reduced at NLO. The PDF
uncertainties are small compared to the scale uncertainties even at NLO and
therefore we will not consider them any further. 
 
Inclusive cross section results for $t\bar{t}Z$ production at the LHC at 8 and
13 TeV for the different operators are shown in Tables~\ref{tab:sigmattZ8} and
\ref{tab:sigmattZ13}. We include the LO and NLO results for $\sigma_i^{(1)}$ and $\sigma_{ii}^{(2)}$, the
corresponding K-factors, the ratio of the dimension-six contribution over the SM
and the ratio of the squared $\mathcal{O}(\Lambda^{-4})$ contributions over the
$\mathcal{O}(\Lambda^{-2})$ one. Statistical uncertainties are not shown unless
they are comparable to the scale uncertainties. The scale uncertainties are
significantly reduced at NLO similarly to the SM predictions. We note that the
ratios over the SM are significantly less sensitive to scale variations
compared to the cross-section numbers. 
 
\begin{table}[h]
\renewcommand{\arraystretch}{1.5}
\footnotesize
\begin{center}
\begin{tabular}{cccccc}
\hline
   8TeV & $\mathcal{O}_{tG}$ &  $\mathcal{O}^{(3)}_{\phi Q}$ &  $\mathcal{O}_{\phi t}$ &  $\mathcal{O}_{tW}$ \\
\hline
 $\sigma_{i,LO}^{(1)}$  & $ 76.1^{ +41.9 \% }_{ -27.1 \% }$ & $ 18.6^{ +45.2 \% }_{ -28.6 \% }$  & $ 12.5^{ +44.6 \% }_{ -28.3 \% }$ & $ 0.077(8)^{ +46.6 \% }_{ -43.2 \% }$\\
  $\sigma_{i,NLO}^{(1)}$ & $ 78.1^{ +4.1 \% }_{ -10.0 \% }$  & $ 20.8^{ +5.6 \% }_{ -11.5 \% }$ & $ 13.5^{ +4.9 \% }_{ -10.7 \% }$ & $ -0.32(2)^{ +39.1 \% }_{ -67.3 \% }$ \\
 K-factor  & 1.03 & 1.12 & 1.08  & -4.2  \\
 $\sigma_{i,LO}^{(2)}$  & $ 39.9^{ +53.6 \% }_{ -31.8 \% } $ & $ 0.73(2)^{ +45.2 \% }_{ -28.8 \% } $ & $ 0.73(2)^{ +46.3 \% }_{ -28.8 \% } $ & $ 4.14^{ +50.1 \% }_{ -30.7 \% } $ \\
 $\sigma_{i,NLO}^{(2)}$  & $ 39.8^{ +4.7 \% }_{ -9.4 \% } $ & $ 0.8(2)^{ +5.4 \% }_{ -9.1 \% } $  & $ 0.8(2)^{ +7.4 \% }_{ -8.3 \% } $ & $ 4.81^{ +6.2 \% }_{ -12.5 \% } $ \\
 $\sigma_{i,LO}^{(1)}/\sigma_{SM,LO}$  & $ 0.368^{ +0.4 \% }_{ -0.4 \% }$ & $ 0.0899^{ +2.7 \% }_{ -2.5 \% }$ & $ 0.0604^{ +2.3 \% }_{ -2.0 \% }$ & $ 0.00037(4)^{ +33.6 \% }_{ -42.5 \% }$  \\
 $\sigma_{i,NLO}^{(1)}/\sigma_{SM,NLO}$  & $ 0.345^{ +1.3 \% }_{ -2.8 \% }$ & $ 0.0918^{ +0.6 \% }_{ -1.0 \% }$ & $ 0.0595^{ +0.8 \% }_{ -2.3 \% }$ & $ -0.0014(1)^{ +31.4 \% }_{ -56.8 \% }$\\
 $\sigma_{i,LO}^{(2)}/\sigma_{i,LO}^{(1)}$ & $ 0.524^{ +8.2 \% }_{ -6.5 \% }$ & $ 0.039(1)^{ +0.3 \% }_{ -0.5 \% }$ & $ 0.058(2)^{ +1.2 \% }_{ -0.7 \% }$ & $ 54(6)^{ +84.7 \% }_{ -29.1 \% }$ \\
 $\sigma_{i,NLO}^{(2)}/\sigma_{i,NLO}^{(1)}$ & $ 0.509^{ +1.4 \% }_{ -8.4 \% }$ & $ 0.037(8)^{ +2.7 \% }_{ -4.5 \% }$ & $ 0.06(1)^{ +3.2 \% }_{ -5.9 \% }$ & $ -15(1)^{ +36.9 \% }_{ -43.5 \% }$ \\
\hline
\end{tabular}
\end{center}
 \caption{\label{tab:sigmattZ8} Cross sections (in fb) for $t\bar{t}Z$ production at the LHC at $\sqrt{s} =  8$~TeV for the different dimension-six operators. Percentages correspond to scale uncertainties. Integration errors are shown in brackets if these are comparable in size to the scale uncertainties. }  
\end{table}

\begin{table}[h]
\renewcommand{\arraystretch}{1.5}
\footnotesize
\begin{center}
\begin{tabular}{cccccc}
\hline
   13TeV &  $\mathcal{O}_{tG}$ & $\mathcal{O}^{(3)}_{\phi Q}$ &  $\mathcal{O}_{\phi t}$ &  $\mathcal{O}_{tW}$ \\
\hline
 $\sigma_{i,LO}^{(1)}$  & $ 286.7^{ +38.2 \% }_{ -25.5 \% }$ & $ 78.3^{ +40.4 \% }_{ -26.6 \% }  $ & $ 51.6^{ +40.1 \% }_{ -26.4 \% }$ & $ -0.20(3)^{ +88.0 \% }_{ -230.0 \% }$  \\
  $\sigma_{i,NLO}^{(1)}$ & $ 310.5^{ +5.4 \% }_{ -9.7 \% }$   & $ 90.6^{ +7.1 \% }_{ -11.0 \% } $ & $ 57.5^{ +5.8 \% }_{ -10.3 \% } $ & $ -1.7(2)^{ +31.3 \% }_{ -49.1 \% }$  \\ 
 K-factor  & 1.08 & 1.16 & 1.11 & 8.5\\
 $\sigma_{i,LO}^{(2)}$  & $ 258.5^{ +49.7 \% }_{ -30.4 \% } $  & $ 2.8(1)^{ +39.7 \% }_{ -26.9 \% } $ & $ 2.9(1)^{ +39.7 \% }_{ -26.7 \% } $ & $ 20.9^{ +44.3 \% }_{ -28.3 \% } $ \\
 $\sigma_{i,NLO}^{(2)}$  & $ 244.5^{ +4.2 \% }_{ -8.1 \% } $ & $ 3.8(3)^{ +13.2 \% }_{ -14.4 \% } $ & $ 3.9(3)^{ +13.8 \% }_{ -14.6 \% } $ & $ 24.2^{ +6.2 \% }_{ -11.2 \% } $  \\
 $\sigma_{i,LO}^{(1)}/\sigma_{SM,LO}$  & $ 0.376^{ +0.3 \% }_{ -0.3 \% }$ & $ 0.103^{ +1.9 \% }_{ -1.8 \% }$ & $ 0.0677^{ +1.7 \% }_{ -1.6 \% } $ & $ -0.00026(4)^{ +89.5 \% }_{ -167.2 \% }$ \\
 $\sigma_{i,NLO}^{(1)}/\sigma_{SM,NLO}$  & $ 0.353^{ +1.3 \% }_{ -2.4 \% }$ & $ 0.103^{ +0.7 \% }_{ -0.8 \% }$ & $ 0.0654^{ +1.1 \% }_{ -2.1 \% }$ & $ -0.0020(2)^{ +22.9 \% }_{ -38.0 \% }$  \\
 $\sigma_{i,LO}^{(2)}/\sigma_{i,LO}^{(1)}$ & $ 0.902^{ +8.4 \% }_{ -6.7 \% }$  & $ 0.036(1)^{ +0.2 \% }_{ -1.1 \% }$ & $ 0.056(2)^{ +0.6 \% }_{ -0.3 \% }$ & $ -104(16)^{ +60.8 \% }_{ -815.2 \% }$\\
 $\sigma_{i,NLO}^{(2)}/\sigma_{i,NLO}^{(1)}$ & $ 0.787^{ +3.3 \% }_{ -12.8 \% }$  & $ 0.042(4)^{ +5.6 \% }_{ -3.9 \% }$ & $ 0.067(6)^{ +7.6 \% }_{ -4.8 \% } $ & $ -14(1)^{ +29.0 \% }_{ -29.1 \% } $ \\
\hline
\end{tabular}
\end{center}
 \caption{\label{tab:sigmattZ13} Cross sections (in fb) for $t\bar{t}Z$ production at the LHC at $\sqrt{s} =  13$~TeV for the different dimension-six operators. Percentages correspond to scale uncertainties. Integration errors are shown in brackets if these are comparable in size to the scale uncertainties.  }  
\end{table}
 
In the tables, we include the $\mathcal{O}^{(3)}_{\phi Q}$ operator but
not $\mathcal{O}^{(1)}_{\phi Q}$.
Results for $\mathcal{O}^{(1)}_{\phi Q}$ differ by a sign at
$\mathcal{O}(\Lambda^{-2})$ and are identical at $\mathcal{O}(\Lambda^{-4})$.\footnote{This is only approximately true at the cross-section level. There is
	a small contribution from the $bbZ$ vertex which spoils the minus
	sign relation between the two operators. The $bbZ$ vertex contributes
	as we are working in the 5-flavour scheme. Nevertheless this
	contribution is in practice numerically negligible and therefore the
	two operators give opposite contributions at $\mathcal{O}(\Lambda^{-2})$.} Similarly at $\mathcal{O}(\Lambda^{-4})$ the contributions of $\mathcal{O}^{(3)}_{\phi Q}$ and $\mathcal{O}_{\phi t}$ are identical. 
This can be traced back to the way the operators modify the $ttZ$ vertex as
shown in Eq.~\ref{ttZvertex}.  Similarly we do not include the results for
$\mathcal{O}_{tB}$, as they can be obtained from those of $\mathcal{O}_{tW}$ by
multiplying by a factor of $-\textrm{tan}^2\theta_w$ (and $\tan^4\theta_w$ for
the squared contributions). 

The largest contribution is given by the chromomagnetic operator  both at 8 and
13 TeV, reaching almost 40\% of the SM. We find that while
$\mathcal{O}^{(3)}_{\phi Q}$ and $\mathcal{O}_{\phi t}$ give contributions of
6-10\% of the SM for $C_i=1$, $\mathcal{O}_{tW}$ and consequently
$\mathcal{O}_{tB}$ give extremely small contributions reaching at most the per
mille level. While the NLO predictions have significantly reduced theoretical uncertainties, we find that the various ratios of cross-sections
considered are generally stable with respect to QCD corrections (apart from
$\mathcal{O}_{tW}$), and also suffer from much smaller scale uncertainties
compared to the cross-sections. This fact can be exploited to extract information
on the Wilson coefficients. The theoretical errors due to neglecting squared operator contributions
$\sigma^{(2)}_{ii}$ are characterised by the last two rows in the table.  The
results indicate that for order one coefficients neglecting squared contributions
is safe for all operators except for $\mathcal{O}_{tB}$ and $\mathcal{O}_{tW}$.
When placing limits, this assessment should be done for the interval of where
the limits are placed.

We note here the extremely small contribution of the $\mathcal{O}_{tW}$
operator, which also leads to larger statistical uncertainties as currently
it is not possible to compute the interference independently of the other two
contributions. In this case the impact of the EFT amplitude squared is much
larger than its interference with the SM. The small size of the interference is
a result of various effects.
The most important reason is that the dipole interaction, $\sigma^{\mu\nu}q_\nu$,
involves the momentum of the $Z$ boson, and leads to a suppression because
the $Z$ tends to be soft in $t\bar t Z$ production at the LHC.  The same is true also for the
$t\bar t\gamma$ production, as we will see.  By crossing $\gamma$ and $g$,
we have explicitly checked that in the process $g\gamma\to t\bar tg$ this suppression
effect becomes an enhancement, as a large momentum for $\gamma$ is guaranteed 
in the initial state. Apart from this, an additional suppression occurs due to
an accidental cancellation between the contributions of the $gg$ and $q\bar q$
channels, as they are similar in size but come with an opposite sign. This
cancellation leads to a final result which is an order of magnitude smaller
than the individual contributions. Finally, an additional reason could be that
the $O_{tW}$ vertex does not produce the $Z$ boson in its longitudinal state,
which is expected to dominate if it has large momentum.

Finally, comparing 8 and 13 TeV we notice a small increase in the K-factors.
The ratios of the $\mathcal{O}(\Lambda^{-2})$ terms over the SM do not
change significantly. For $\mathcal{O}_{tG}$ we notice a significant increase
of the ratio $\mathcal{O}(\Lambda^{-4})$ over $\mathcal{O}(\Lambda^{-2})$ as
the $\mathcal{O}(\Lambda^{-4})$ contribution grows rapidly with energy, as will be evident also in the differential distributions. 

The corresponding $t\bar{t}\gamma$ results are shown in Table
\ref{tab:sigmattA}. In this case a minimum cut of 20 GeV is
set on the transverse momentum of the photon. We note that here only
three operators contribute: $\mathcal{O}_{tG}$, $\mathcal{O}_{tW}$ and
$\mathcal{O}_{tB}$. For this process, $\mathcal{O}_{tW}$ and $\mathcal{O}_{tB}$ are
indistinguishable and therefore only $\mathcal{O}_{tB}$ is included in the
Table.  The K-factors in this process are larger than those of $t\bar{t}Z$, reaching
1.3 for the SM and $\mathcal{O}_{tG}$ operator but lower for
$\mathcal{O}_{tB}$.  This is due to the soft and collinear
configurations between the photon and the additional jet at NLO, which
however cannot happen if the photon is emitted from an $\mathcal{O}_{tB}$
vertex.

\begin{table}[h]
\renewcommand{\arraystretch}{1.5}
\small
\begin{center}
\begin{tabular}{cccccc}
\hline
   8TeV&  $\mathcal{O}_{tG}$ & $\mathcal{O}_{tB}$ & 13TeV &  $\mathcal{O}_{tG}$ & $\mathcal{O}_{tB}$ \\
\hline
 $\sigma_{i,LO}^{(1)}$   & $ 171.5^{ +38.6 \% }_{ -25.6 \% }$ & $ 5.36^{ +41.8 \% }_{ -27.2 \% }$  &  & $ 564.6^{ +35.4 \% }_{ -24.1 \% }$ & $ 19.5^{ +36.7 \% }_{ -24.9 \% }$  \\
  $\sigma_{i,NLO}^{(1)}$ & $ 218.9^{ +13.3 \% }_{ -13.6 \% }$ & $ 5.85^{ +5.9 \% }_{ -9.9 \% }$ &  & $ 765^{ +14.0 \% }_{ -13.4 \% } $ & $ 19.6^{ +4.3 \% }_{ -6.9 \% } $   \\
 K-factor  & 1.28 & 1.09 &  & 1.35 & 1.01 \\
  $\sigma_{i,LO}^{(2)}$  & $ 29.8^{ +43.5 \% }_{ -27.8 \% } $ & $ 1.98^{ +47.5 \% }_{ -29.6 \% } $ &  & $ 120.6^{ +39.8 \% }_{ -26.2 \% } $  & $ 9.14^{ +42.3 \% }_{ -27.4 \% } $ \\
    $\sigma_{i,NLO}^{(2)}$  & $ 39.2^{ +13.1 \% }_{ -14.4 \% } $ & $ 2.36^{ +7.0 \% }_{ -12.6 \% } $ &  & $ 160.4^{ +12.6 \% }_{ -13.5 \% } $ & $ 10.7^{ +6.7 \% }_{ -11.2 \% } $  \\
 $\sigma_{i,LO}^{(1)}/\sigma_{SM,LO}$ & $ 0.284^{ +0.04 \% }_{ -0.1 \% }$ & $ 0.00888^{ +2.3 \% }_{ -2.2 \% }$ & & $ 0.283^{ +0.1 \% }_{ -0.1 \% }$ & $ 0.00973^{ +0.9 \% }_{ -1.0 \% }$  \\
 $\sigma_{i,NLO}^{(1)}/\sigma_{SM,NLO}$ & $ 0.282^{ +0.13 \% }_{ -0.2 \% }$ & $ 0.0075(1)^{ +4.4 \% }_{ -8.8 \% }$  & & $ 0.281^{ +0.1 \% }_{ -0.1 \% }$ & $ 0.0072(1)^{ +7.5 \% }_{ -11.9 \% }$ \\
 $\sigma_{i,LO}^{(2)}/\sigma_{i,LO}^{(1)}$& $ 0.174^{ +3.5 \% }_{ -3.0 \% }$ & $ 0.370^{ +4.0 \% }_{ -3.3 \% }$  &  & $ 0.214^{ +3.3 \% }_{ -2.8 \% }$ & $ 0.470^{ +4.1 \% }_{ -3.4 \% }$ \\
 $\sigma_{i,NLO}^{(2)}/\sigma_{i,NLO}^{(1)}$ & $ 0.179^{ +0.5 \% }_{ -0.9 \% }$  &$ 0.404(7)^{ +3.5 \% }_{ -3.0 \% }$ & & $ 0.201^{ +1.1 \% }_{ -1.3 \% }$ & $ 0.55(1)^{ +6.1 \% }_{ -4.6 \% }$ \\
\hline
\end{tabular}
\end{center}
 \caption{\label{tab:sigmattA} Cross sections (in fb) for $t\bar{t}\gamma$ production at the LHC at $\sqrt{s} =  8$~TeV and $\sqrt{s} =  13$~TeV for the different dimension-six operators. Percentages correspond to scale uncertainties. Integration errors are shown in brackets if these are comparable in size to the scale uncertainties. A $p_T(\gamma)>20$ GeV cut is imposed. }  
\end{table}

Similar conclusions to the $t\bar{t}Z$ can be drawn for $t \bar t \gamma$
regarding the operator contributions.  The chromomagnetic operator contributes
the most. Neglecting squared contributions is safe for $C_i\lesssim1$, at both 8 and 13~TeV,
but starts to become questionable (and therefore the corresponding uncertainty is increased) 
as the coefficients reach order of a few, with the relative contribution of $\sigma^{(2)}_{ii}$ increasing
from 8 to 13 TeV. The contribution of the $\mathcal{O}_{tW}$ and
$\mathcal{O}_{tB}$ operators are 1\% of the SM and significantly smaller than
the $\mathcal{O}_{tG}$ one. While the $\mathcal{O}_{tW}$ and $\mathcal{O}_{tG}$
operators lead to the same structure in the $tt\gamma$ and $ttg$ vertices
respectively, similar to $ttZ$ production, the effect of $\mathcal{O}_{tW}$ on
the $gg\to t\bar{t}\gamma$ amplitude at typical LHC energies is suppressed
compared with that of $\mathcal{O}_{tG}$. By examining the crossed amplitude,
$g\gamma \to t\bar{t}g$, we see that the two operators give contributions of
the same order, as they both enter in the production side of the process and
more momentum passes through the EFT vertices. We also note here that the
K-factors for the operators are not the same as those as for the SM
contribution which implies that combining the SM K-factor and LO EFT
predictions does not provide an accurate prediction for the EFT contribution at
NLO in QCD. 

We next examine $t \bar t l^+ l^-$. For an invariant mass of the lepton pair around the $Z$ mass, this process is dominated by $t  \bar t Z$ with leptonically decaying $Z$, the mode that the ATLAS and CMS experiments at the LHC are most sensitive to. Generally it also includes the contribution of $t \bar t \gamma^*$. As the EFT operators we study do not enter the vertices connected to leptons, we restrict our attention to $t \bar t \mu^+ \mu^-$ \footnote{We note here that a contribution from 4-fermion operators describing the $t\bar{t}\mu^+ \mu^-$ interaction enter in this process in the off-peak regions. As the main contribution comes from the $Z-$peak we postpone the study of these operators to future work.}.
We collect the results for $t\bar{t}\mu^+\mu^-$ at LO and NLO at 8 and 13~TeV in Tables \ref{tab:sigmattmumu8} and \ref{tab:sigmattmumu13}. In this case the photon and $Z$ amplitudes and their interference is included. For the $t\bar{t}\mu^+\mu^-$ results, the scale and PDF choices are identical to those for the inclusive $t\bar{t}Z/\gamma$ processes. A lower cut of 10~GeV is imposed on the invariant mass of the lepton pair. No other cuts are imposed on the leptons. 

\begin{table}[h]
\renewcommand{\arraystretch}{1.5}
\scriptsize
\begin{center}
\begin{tabular}{cccccccc}
\hline
   8TeV &  $\mathcal{O}_{tG}$ & $\mathcal{O}^{(3)}_{\phi Q}$ &  $\mathcal{O}_{\phi t}$ & $\mathcal{O}_{tB}$ &  $\mathcal{O}_{tW}$ \\
\hline
 $\sigma_{i,LO}^{(1)}$  & $ 3.07^{ +41.5 \% }_{ -26.9 \% }$ & $ 0.613^{ +45.2 \% }_{ -28.6 \% }$ & $ 0.413^{ +44.6 \% }_{ -28.3 \% }$ & $ 0.0101^{ +43.2 \% }_{ -27.6 \% }$ & $ 0.0121(6)^{ +29.2 \% }_{ -21.5 \% }$ \\
  $\sigma_{i,NLO}^{(1)}$ & $ 3.21^{ +5.1 \% }_{ -10.4 \% }$ & $ 0.683^{ +5.4 \% }_{ -11.3 \% } $ & $ 0.447^{ +4.8 \% }_{ -10.9 \% }$  & $ 0.012(1)^{ +8.9 \% }_{ -12.2 \% }$ & $ -0.003(2)^{ +113.9 \% }_{ -205.9 \% }$ \\
 K-factor  & 1.05 & 1.11 &1.08 & 1.2 & -0.3 \\
 $\sigma_{i,LO}^{(2)}$  & $ 1.42^{ +52.9 \% }_{ -31.6 \% } $ & $ 0.0238^{ +45.2 \% }_{ -28.7 \% } $ & $ 0.0234^{ +45.8 \% }_{ -28.7 \% } $ & $ 0.0213^{ +49.8 \% }_{ -30.6 \% } $  & $ 0.147^{ +50.1 \% }_{ -30.7 \% } $ \\
  $\sigma_{i,NLO}^{(2)}$  & $ 1.41^{ +4.5 \% }_{ -9.7 \% } $  & $ 0.0275^{ +6.4 \% }_{ -11.7 \% } $ & $ 0.0259^{ +5.0 \% }_{ -11.4 \% } $  & $ 0.0249^{ +6.5 \% }_{ -12.6 \% } $ & $ 0.171^{ +6.3 \% }_{ -12.5 \% } $  \\
 $\sigma_{i,LO}^{(1)}/\sigma_{SM,LO}$  & $ 0.350^{ +0.4 \% }_{ -0.4 \% }$  & $ 0.0698^{ +3.1 \% }_{ -2.8 \% }$ & $ 0.0470^{ +2.6 \% }_{ -2.3 \% }$ & $ 0.00115^{ +1.6 \% }_{ -1.7 \% } $  & $ 0.0014(1)^{ +6.9 \% }_{ -8.4 \% }$ \\
 $\sigma_{i,NLO}^{(1)}/\sigma_{SM,NLO}$  & $ 0.327^{ +1.2 \% }_{ -2.4 \% } $ & $ 0.0695^{ +1.0 \% }_{ -2.3 \% }$ & $ 0.0455^{ +1.3 \% }_{ -2.8 \% }$ & $ 0.0012(1)^{ +2.0 \% }_{ -1.5 \% }$  & $ -0.0004(2)^{ +115.7 \% }_{ -184.1 \% }$  \\
 $\sigma_{i,LO}^{(2)}/\sigma_{i,LO}^{(1)}$ & $ 0.461^{ +8.1 \% }_{ -6.5 \% }$ & $ 0.0388^{ +0.0 \% }_{ -0.1 \% }$ & $ 0.0567^{ +0.8 \% }_{ -0.7 \% }$ & $ 2.11(5)^{ +5.2 \% }_{ -4.1 \% } $ & $ 12.2(6)^{ +16.3 \% }_{ -11.7 \% } $ \\
 $\sigma_{i,NLO}^{(2)}/\sigma_{i,NLO}^{(1)}$ & $ 0.440^{ +1.7 \% }_{ -8.3 \% }$ & $ 0.0403(8)^{ +1.0 \% }_{ -0.7 \% }$ & $ 0.058(2)^{ +0.4 \% }_{ -0.6 \% }$ & $ 2.1(1)^{ +2.6 \% }_{ -2.8 \% } $  & $ -49(23)^{ +730.1 \% }_{ -332.3 \% }$ \\
\hline
\end{tabular}
\end{center}
 \caption{\label{tab:sigmattmumu8} Cross sections (in fb) for $t\bar{t}\mu^+\mu^-$ production at the LHC at $\sqrt{s} =  8$~TeV for the different dimension-six operators.  A $m(\ell \ell)>10$ GeV cut is applied to the lepton pair. Percentages correspond to scale uncertainties. Integration errors are shown in brackets if these are comparable in size to the scale uncertainties.}  
\end{table}

\begin{table}[h]
\renewcommand{\arraystretch}{1.5}
\scriptsize
\begin{center}
\begin{tabular}{ccccccc}
\hline
   13TeV & $\mathcal{O}_{tG}$ &  $\mathcal{O}^{(3)}_{\phi Q}$ &  $\mathcal{O}_{\phi t}$ & $\mathcal{O}_{tB}$ &  $\mathcal{O}_{tW}$ \\
\hline
 $\sigma_{i,LO}^{(1)}$  & $ 11.28^{ +37.8 \% }_{ -25.2 \% }$ & $ 2.584^{ +40.4 \% }_{ -26.6 \% }$ & $ 1.701^{ +40.1 \% }_{ -26.4 \% }$ & $ 0.034(1)^{ +36.9 \% }_{ -25.1 \% } $  & $ 0.025(3)^{ +29.4 \% }_{ -24.8 \% }$ \\ 
  $\sigma_{i,NLO}^{(1)}$ & $ 12.57^{ +6.7 \% }_{ -10.3 \% }$ & $ 2.976^{ +6.7 \% }_{ -10.8 \% }$ & $ 1.891^{ +5.4 \% }_{ -10.1 \% }$  & $ 0.046(2)^{ +13.0 \% }_{ -12.7 \% } $  & $ -0.042(9)^{ +44.6 \% }_{ -73.2 \% }$ \\  
 K-factor  & 1.11 & 1.15 & 1.11  & 1.3 & -1.7 \\
  $\sigma_{i,LO}^{(2)}$  & $ 8.957^{ +49.3 \% }_{ -30.2 \% } $  & $ 0.101^{ +40.4 \% }_{ -26.6 \% } $  & $ 0.0998^{ +40.6 \% }_{ -26.6 \% } $ & $ 0.1073^{ +44.3 \% }_{ -28.3 \% } $  & $ 0.745^{ +44.4 \% }_{ -28.4 \% } $ \\
  $\sigma_{i,NLO}^{(2)}$  & $ 8.49^{ +4.1 \% }_{ -7.4 \% } $ & $ 0.1168^{ +7.1 \% }_{ -11.0 \% } $  & $ 0.112(3)^{ +5.5 \% }_{ -10.0 \% } $ & $ 0.1231^{ +6.2 \% }_{ -11.0 \% } $  & $ 0.851^{ +5.9 \% }_{ -11.0 \% } $ \\
 $\sigma_{i,LO}^{(1)}/\sigma_{SM,LO}$  & $ 0.356^{ +0.3 \% }_{ -0.2 \% }$ & $ 0.0816^{ +2.2 \% }_{ -2.0 \% }$ & $ 0.0537^{ +2.0 \% }_{ -1.8 \% } $ & $ 0.00108(3)^{ +0.3 \% }_{ -0.5 \% } $ & $ 0.0008(1)^{ +12.7 \% }_{ -16.1 \% } $  \\
 $\sigma_{i,NLO}^{(1)}/\sigma_{SM,NLO}$  & $ 0.335^{ +1.2 \% }_{ -2.2 \% } $ & $ 0.0793^{ +1.1 \% }_{ -2.2 \% } $ & $ 0.0504^{ +1.5 \% }_{ -3.5 \% } $ & $ 0.0012(1)^{ +3.6 \% }_{ -1.6 \% } $ & $ -0.0011(2)^{ +37.6 \% }_{ -58.7 \% } $  \\
 $\sigma_{i,LO}^{(2)}/\sigma_{i,LO}^{(1)}$ & $ 0.794^{ +8.4 \% }_{ -6.7 \% }$ & $ 0.0390^{ +0.03 \% }_{ -0.02 \% } $ & $ 0.0586^{ +0.5 \% }_{ -0.4 \% } $ & $ 3.15(9)^{ +5.5 \% }_{ -4.6 \% } $  & $ 29(4)^{ +25.3 \% }_{ -15.2 \% } $ \\ 
 $\sigma_{i,NLO}^{(2)}/\sigma_{i,NLO}^{(1)}$ & $ 0.676^{ +3.6 \% }_{ -13.3 \% } $ & $ 0.0393^{ +0.3 \% }_{ -0.2 \% } $ & $ 0.059(1)^{ +0.2 \% }_{ -0.2 \% } $ & $ 2.7(1)^{ +2.1 \% }_{ -6.8 \% } $  & $ -20(5)^{ +39.2 \% }_{ -60.7 \% } $ \\
\hline
\end{tabular}
\end{center}
 \caption{\label{tab:sigmattmumu13} Cross sections (in fb) for $t\bar{t}\mu^+\mu^-$ production at the LHC at $\sqrt{s} =  13$~TeV  for the different dimension-six operators.  A $m(\ell \ell)>10$ GeV cut is applied to the lepton pair. Percentages correspond to scale uncertainties. Integration errors are shown in brackets if these are comparable in size to the scale uncertainties.}  
\end{table}

All six operators contribute to this process. Results for $\mathcal{O}^{(1)}_{\phi Q}$ differ by a sign at $\mathcal{O}(\Lambda^{-2})$ and are identical at $\mathcal{O}(\Lambda^{-4})$ to those of  $\mathcal{O}^{(3)}_{\phi Q}$ and therefore we show only one of the two. The cross-section is dominated by the region close to the $Z-$mass peak and therefore the K-factors and relative contributions of the operators are similar to those of the $t\bar{t}Z$ process. The chromomagnetic operator contributes at the 35\% level, while the other three current operators give a contribution at the 4-7\% level.
 
The contributions of  $\mathcal{O}_{tW}$ and $\mathcal{O}_{tB}$ at $\mathcal{O}(\Lambda^{-2})$ are at the per mille level and subdominant compared to the $\mathcal{O}(\Lambda^{-4})$ contributions. Effectively this means that with our method of extracting the interference contribution we are always very limited statistically. Even maximising the interference contribution by choosing the appropriate value of the coefficient is not enough to give us good statistics, in particular at NLO which is evident in the quoted statistical uncertainties. 

\begin{figure}[b!]
 \begin{minipage}[t]{0.5\linewidth}
\centering
\includegraphics[trim=1cm 0 0 0, scale =1.2]{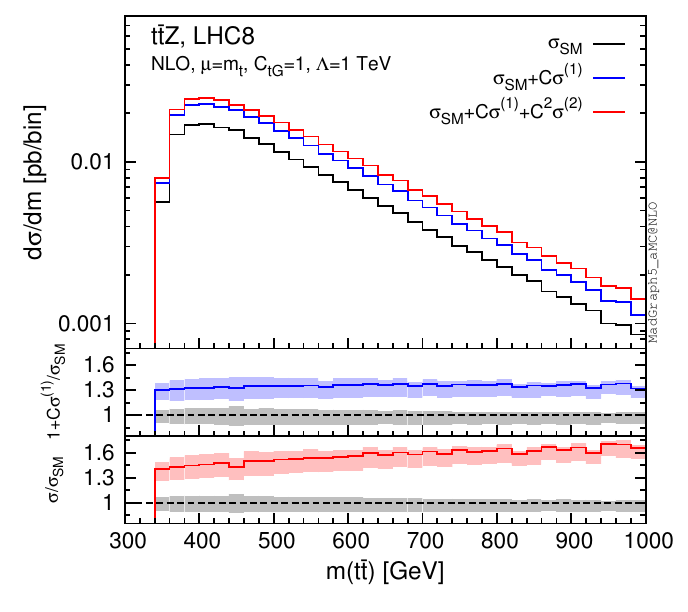}
\end{minipage}
\hspace{0.5cm}
 \begin{minipage}[t]{0.5\linewidth}
 \centering
 \includegraphics[trim=1cm 0 0 0,scale=1.2]{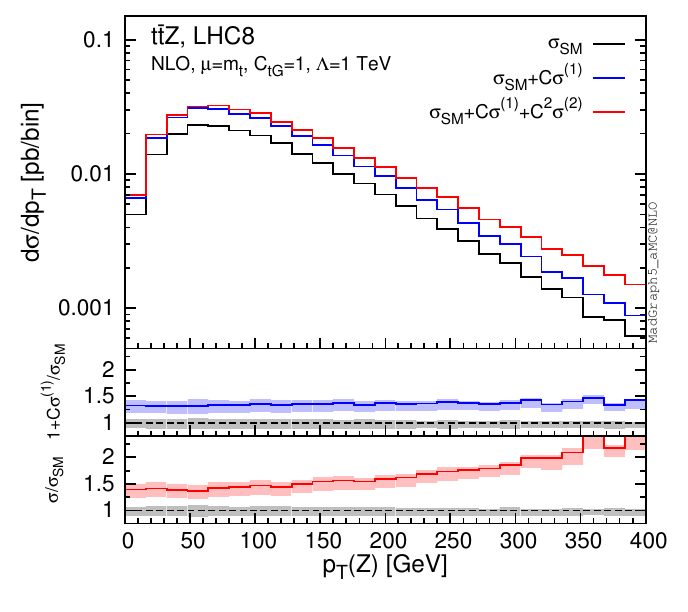}
 \end{minipage}
  \begin{minipage}[t]{0.5\linewidth}
\centering
\includegraphics[trim=1cm 0 0 0, scale =1.2]{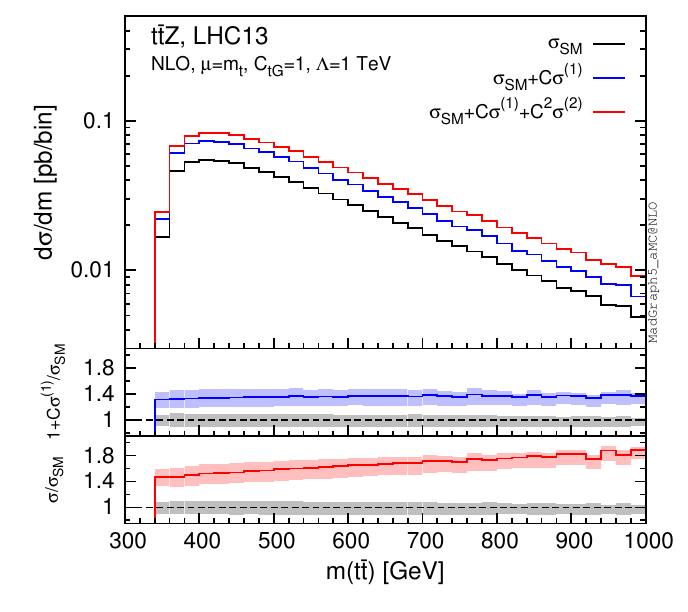}
\end{minipage}
\hspace{0.5cm}
 \begin{minipage}[t]{0.5\linewidth}
 \centering
 \includegraphics[trim=1cm 0 0 0,scale=1.2]{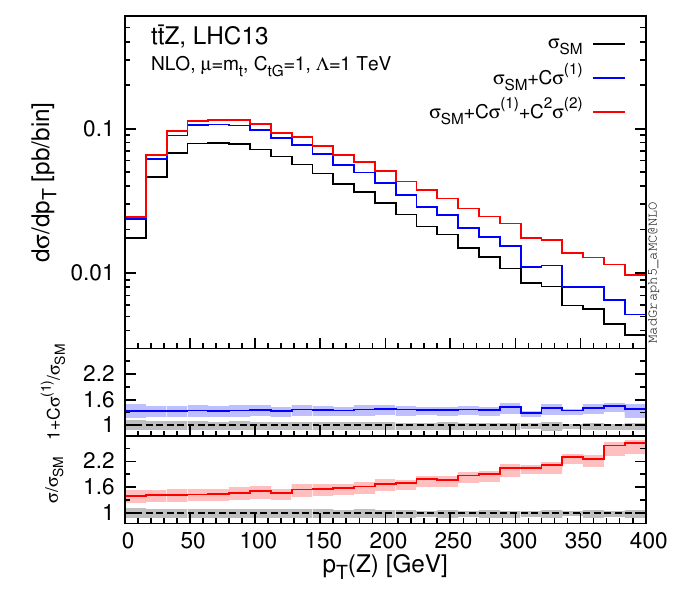}
 \end{minipage}
\caption{\label{fig:ttz_8_13_tg} Invariant mass distributions for the top quark pair and $Z$ $p_T$ distribution at 8 and 13 TeV for the chromomagnetic operator for $C_{tG}=1$ and $\Lambda=1$~TeV. Scale uncertainty bands are shown.} 
\end{figure}

\subsection{Differential distributions}

Differential distributions are obtained at NLO for the $pp\to t\bar{t}Z$,
$pp\to t\bar{t}\gamma $ and  $pp \rightarrow t \bar t \mu^+ \mu^-$ processes.
This can be done also at NLO with matching to the PS,
and with top quarks decayed keeping spin correlations \cite{Artoisenet:2012st},
all implemented in the
{\sc MG5\_aMC} framework. 

\begin{figure}[b!]
 \begin{minipage}[t]{0.5\linewidth}
\centering
\includegraphics[trim=1cm 0 0 0, scale =1.2]{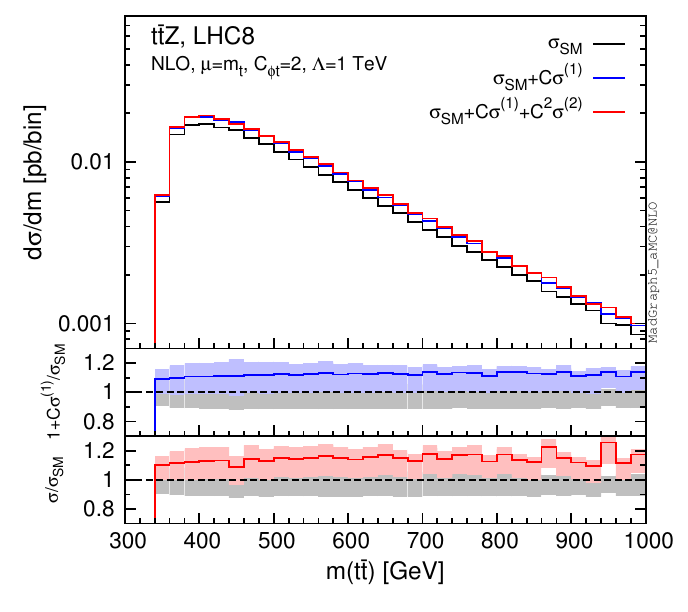}
\label{mttx_tg_8}
\end{minipage}
\hspace{0.5cm}
 \begin{minipage}[t]{0.5\linewidth}
 \centering
 \includegraphics[trim=1cm 0 0 0,scale=1.2]{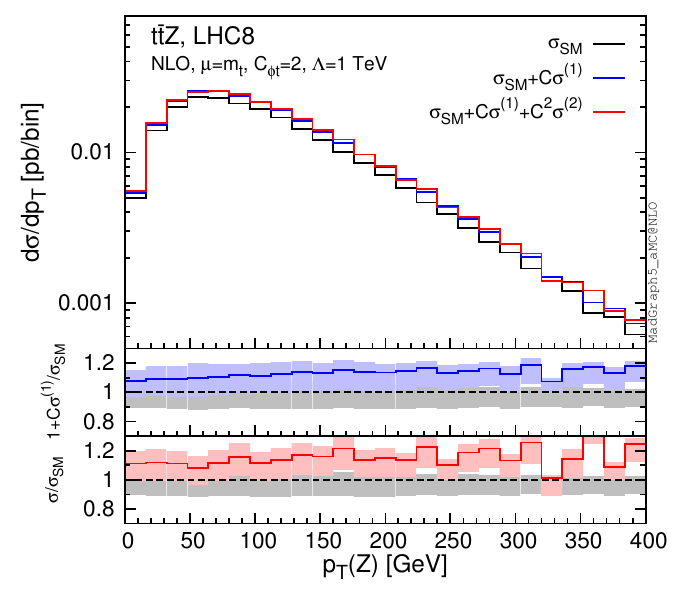}
 \end{minipage}
 \begin{minipage}[t]{0.5\linewidth}
\centering
\includegraphics[trim=1cm 0 0 0, scale =1.2]{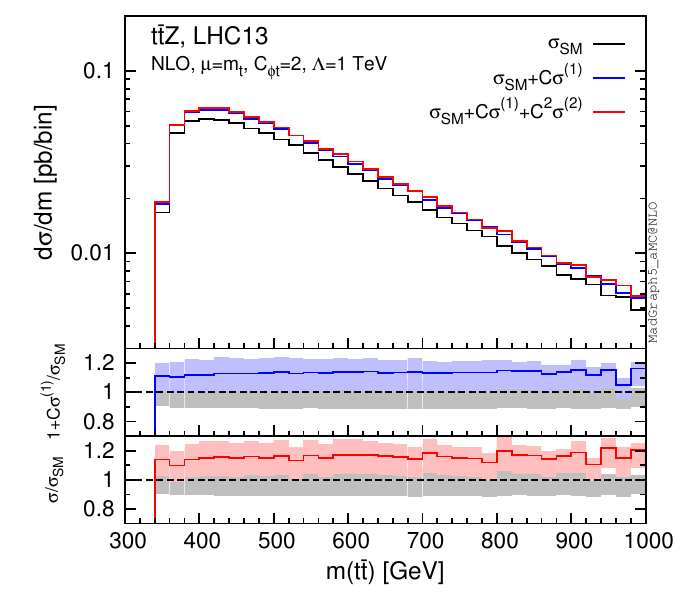}
\label{mttx_tg_8}
\end{minipage}
\hspace{0.5cm}
 \begin{minipage}[t]{0.5\linewidth}
 \centering
 \includegraphics[trim=1cm 0 0 0,scale=1.2]{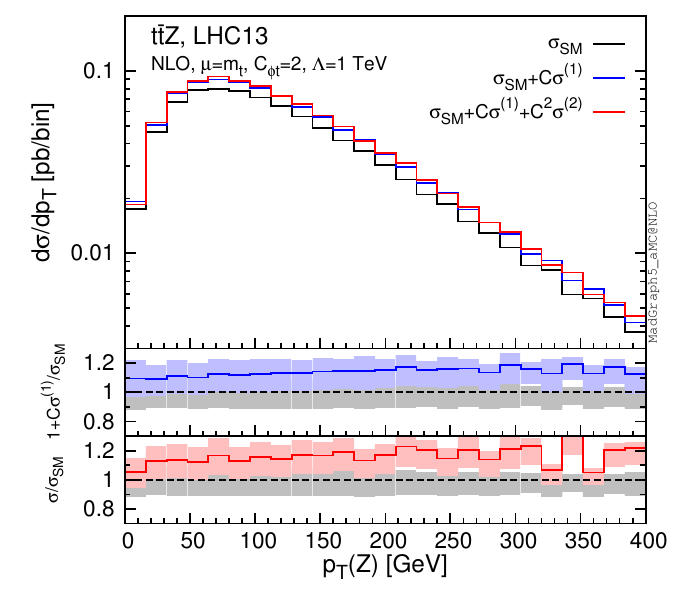}
 \end{minipage}
\caption{\label{fig:ttz_8_13_ft} Invariant mass distributions for the top quark pair and $Z$ $p_T$ distribution at 8 and 13 TeV for the $\mathcal{O}_{\phi t}$ operator for $C_{\phi t}=2$ and $\Lambda=1$~TeV.  Scale uncertainty bands are shown.} 
\end{figure}
\noindent
Hence our approach can be used directly in
realistic experimental simulation, with NLO+PS event generation, which
allows for more detailed studies of possible EFT signals.
In this work, for illustration purpose, we keep results simple by only
presenting fixed order NLO distributions.  No kinematical cuts are applied
except for the $m_(\ell \ell)>10$ GeV and $p_T(\gamma)>20$ GeV generation cuts.  We show
results obtained with one non-zero operator coefficient at a time, with $\Lambda=1$ TeV,
and SM results are given for comparison.

We start by showing the distributions obtained for the $\mathcal{O}_{tG}$
operator at 8 and 13 TeV in Fig.~\ref{fig:ttz_8_13_tg}. We show as a reference the invariant
mass distribution of the top quark pair and the transverse momentum of the $Z$.
In the plots we show the SM prediction $\sigma_{\mathrm{SM}}$, the result for $C_{tG}=1,\, \Lambda=1$
TeV i) adding only the interference $\sigma^{(1)}_{i}$ and ii) adding both the interference and the
squared terms $\sigma^{(2)}_{ii}$.

\begin{figure}[b!]
 \begin{minipage}[t]{0.5\linewidth}
\centering
\includegraphics[trim=1cm 0 0 0, scale =1.2]{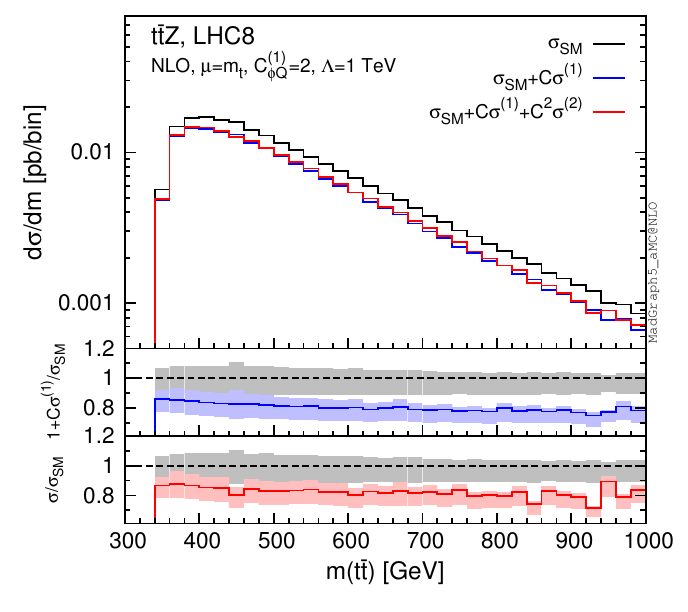}
\label{mttx_tg_8}
\end{minipage}
\hspace{0.5cm}
 \begin{minipage}[t]{0.5\linewidth}
 \centering
 \includegraphics[trim=1cm 0 0 0,scale=1.2]{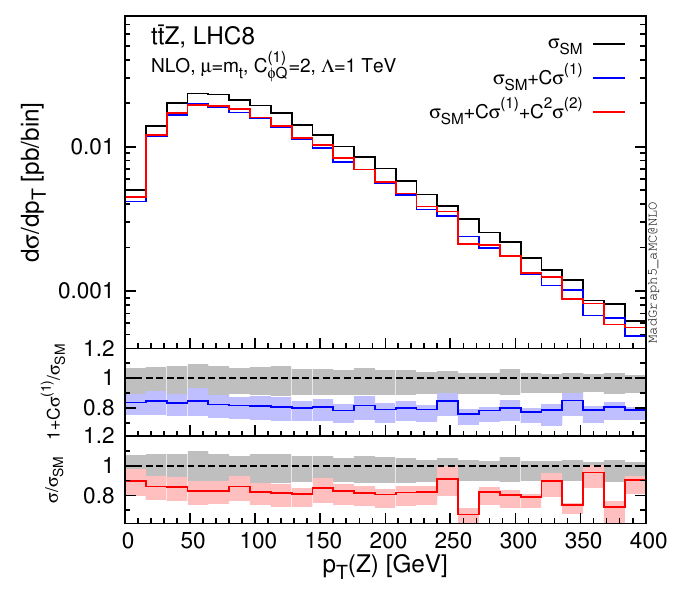}
 \end{minipage}
 \begin{minipage}[t]{0.5\linewidth}
\centering
\includegraphics[trim=1cm 0 0 0, scale =1.2]{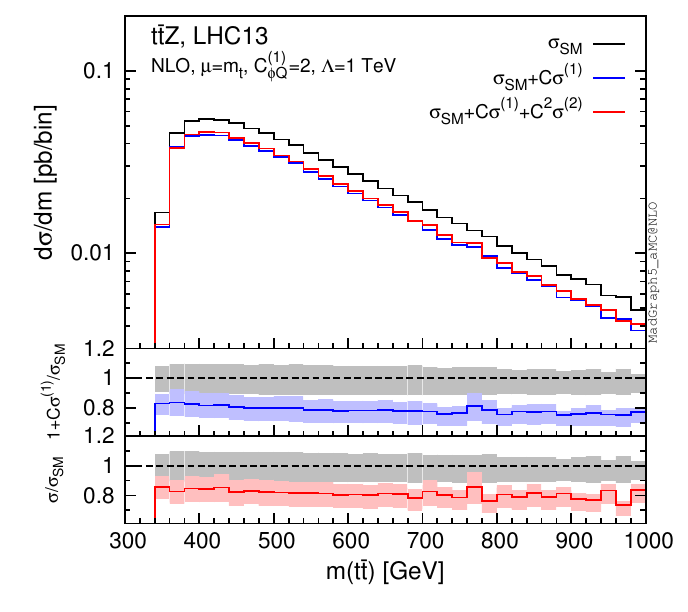}
\label{mttx_tg_8}
\end{minipage}
\hspace{0.5cm}
 \begin{minipage}[t]{0.5\linewidth}
 \centering
 \includegraphics[trim=1cm 0 0 0,scale=1.2]{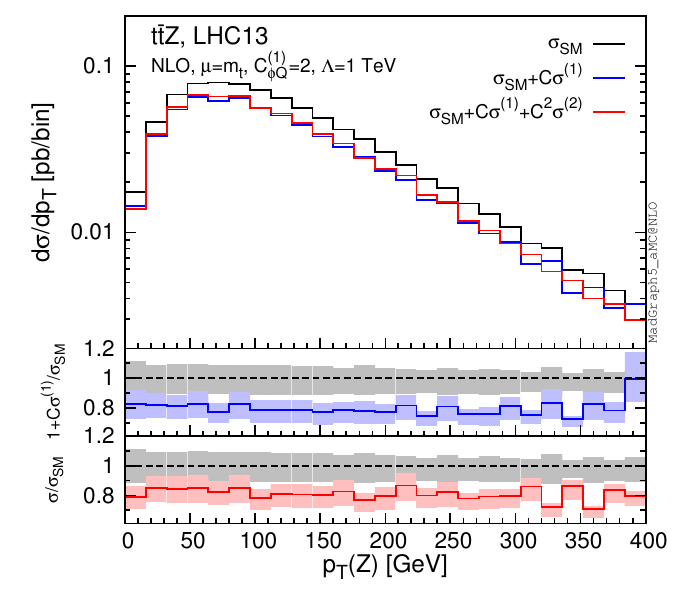}
 \end{minipage}
\caption{\label{fig:ttz_8_13_fQ1} Invariant mass distributions for the top quark pair and $Z$ $p_T$ distribution at 8 and 13 TeV for the $\mathcal{O}_{\phi Q}^{(1)}$ operator for $C_{\phi Q^{(1)}}=2$ and $\Lambda=1$~TeV.  Scale uncertainty bands are shown.  } 
\end{figure}
\noindent
We also include the corresponding ratios over the SM prediction
and the scale uncertainty bands. It is clear that while the interference
contribution is not changing the distribution shape, the
$\mathcal{O}(\Lambda^{-4})$ contribution is growing fast at high energies with
the effect being more evident at 13~TeV in both distributions shown here. Similar observations can be made
 for other observables, such as the transverse momentum of the top. 
 
\begin{figure}[b!]
 \begin{minipage}[t]{0.5\linewidth}
\centering
\includegraphics[trim=1cm 0 0 0, scale =1.2]{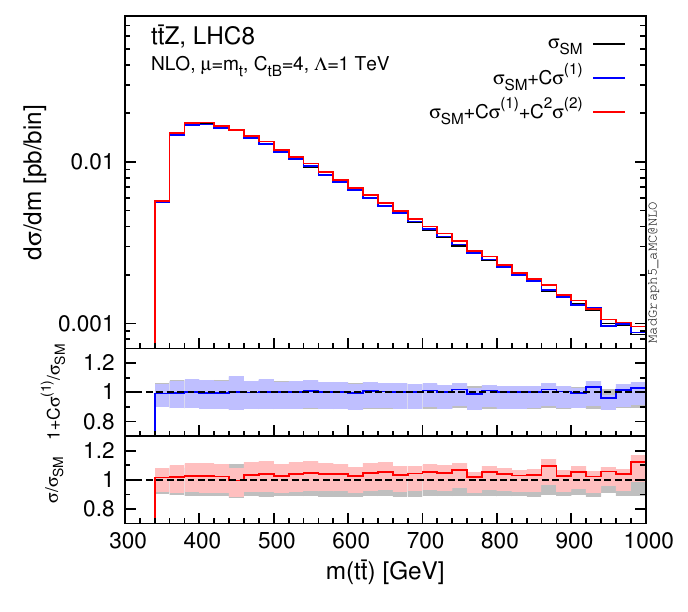}
\label{mttx_tg_8}
\end{minipage}
\hspace{0.5cm}
 \begin{minipage}[t]{0.5\linewidth}
 \centering
 \includegraphics[trim=1cm 0 0 0,scale=1.2]{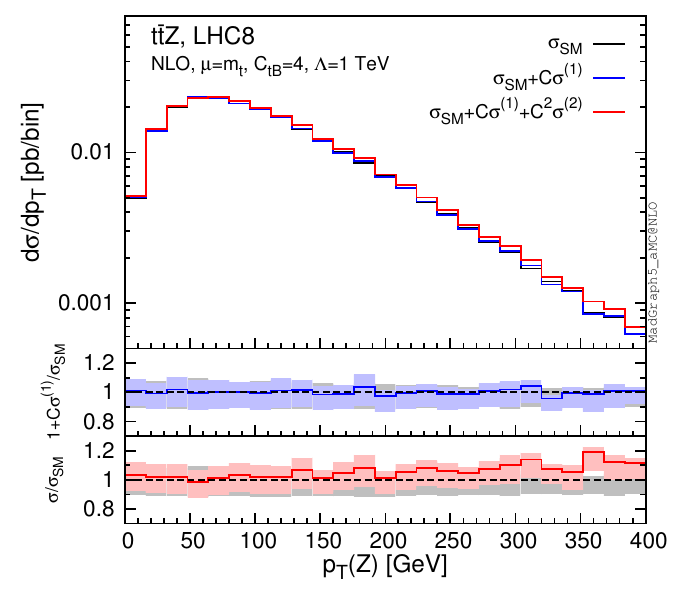}
 \end{minipage}
 \begin{minipage}[t]{0.5\linewidth}
\centering
\includegraphics[trim=1cm 0 0 0, scale =1.2]{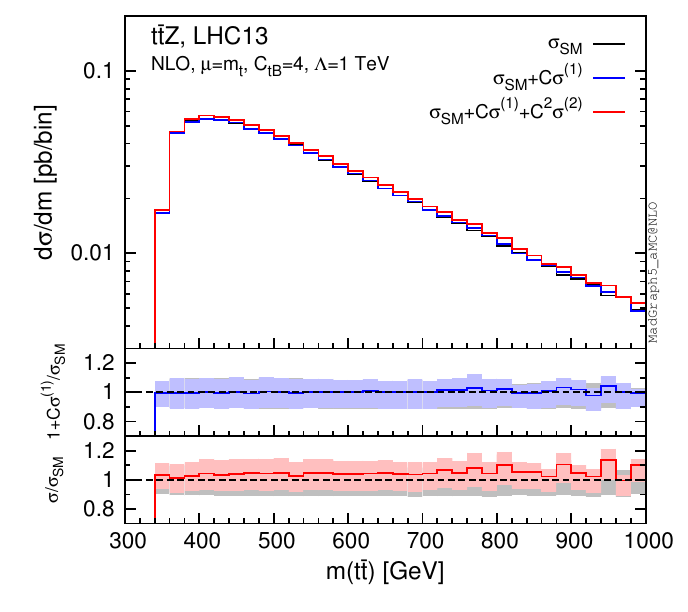}
\label{mttx_tg_8}
\end{minipage}
\hspace{0.5cm}
 \begin{minipage}[t]{0.5\linewidth}
 \centering
 \includegraphics[trim=1cm 0 0 0,scale=1.2]{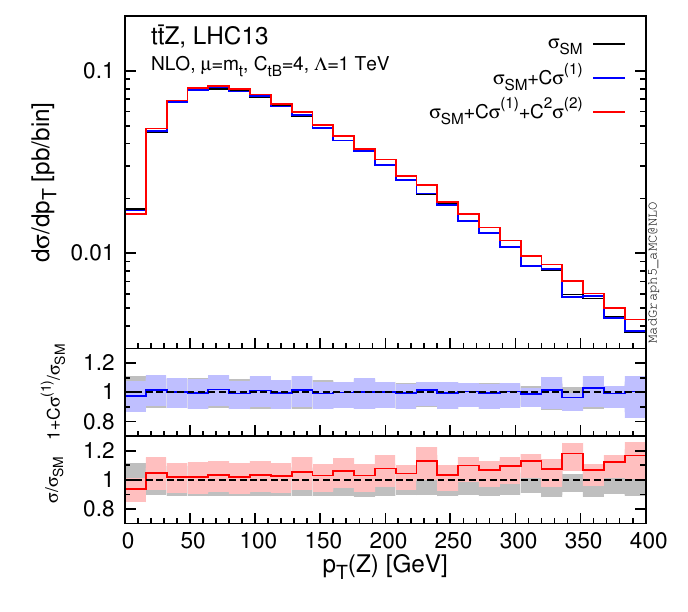}
 \end{minipage}
\caption{\label{fig:ttz_8_13_tB} Invariant mass distributions for the top quark pair and $Z$ $p_T$ distribution at 8 and 13 TeV for the $\mathcal{O}_{t B}$ operator for $C_{tB}=4$ and $\Lambda=1$~TeV.  Scale uncertainty bands are shown.  } 
\end{figure}

Results for the $\mathcal{O}_{\phi t}$ and $\mathcal{O}_{\phi Q}^{(1)}$ are
shown in Fig.~\ref{fig:ttz_8_13_ft} and \ref{fig:ttz_8_13_fQ1} respectively. In this case we set the Wilson coefficients to 2, in 
order to obtain visible deviations from the SM. These values are allowed by the current constraints. 
For these operators the
$\mathcal{O}(\Lambda^{-4})$ contribution is significantly smaller than the
$\mathcal{O}(\Lambda^{-2})$ and does not significantly alter the shape of the differential
distributions as seen in the flat ratios for both the $t\bar{t}$ invariant mass
and $Z$ $p_T$ distributions. Results for $\mathcal{O}_{\phi Q}^{(3)}$ are
identical to those of $\mathcal{O}_{\phi Q}^{(1)}$ (with a minus sign), so we do not include them
for brevity. 

\begin{figure}[b!]
 \begin{minipage}[t]{0.5\linewidth}
\centering
\includegraphics[trim=1cm 0 0 0, scale =1.2]{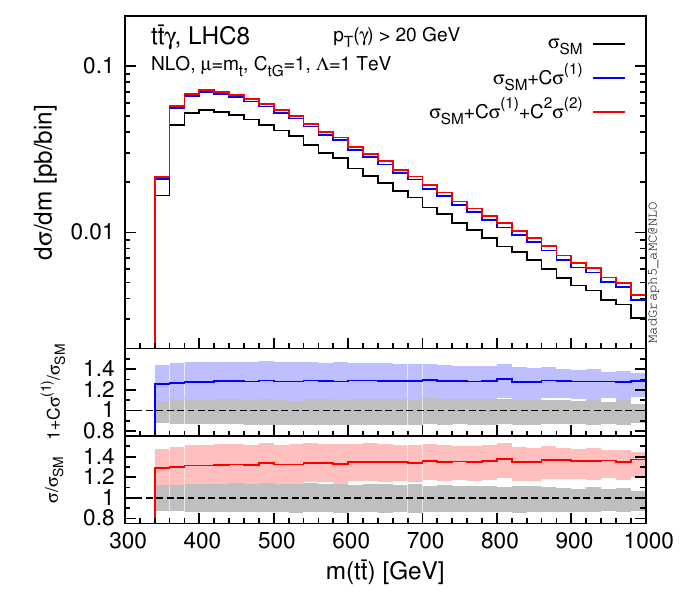}
\label{mttx_tg_8}
\end{minipage}
\hspace{0.5cm}
 \begin{minipage}[t]{0.5\linewidth}
 \centering
 \includegraphics[trim=1cm 0 0 0,scale=1.2]{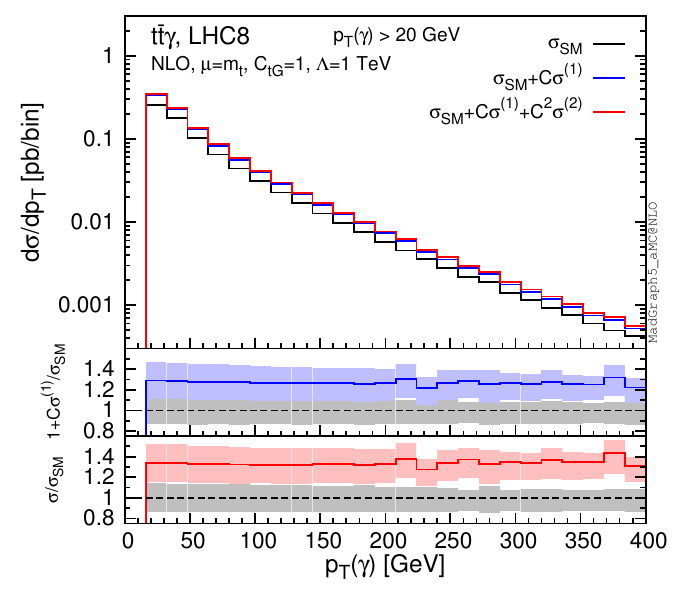}
 \end{minipage}
 \begin{minipage}[t]{0.5\linewidth}
\centering
\includegraphics[trim=1cm 0 0 0, scale =1.2]{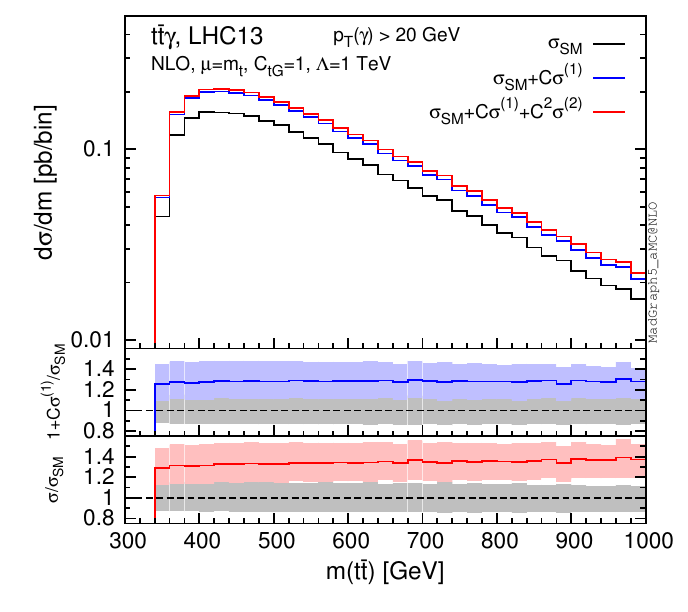}
\label{mttx_tg_8}
\end{minipage}
\hspace{0.5cm}
 \begin{minipage}[t]{0.5\linewidth}
 \centering
 \includegraphics[trim=1cm 0 0 0,scale=1.2]{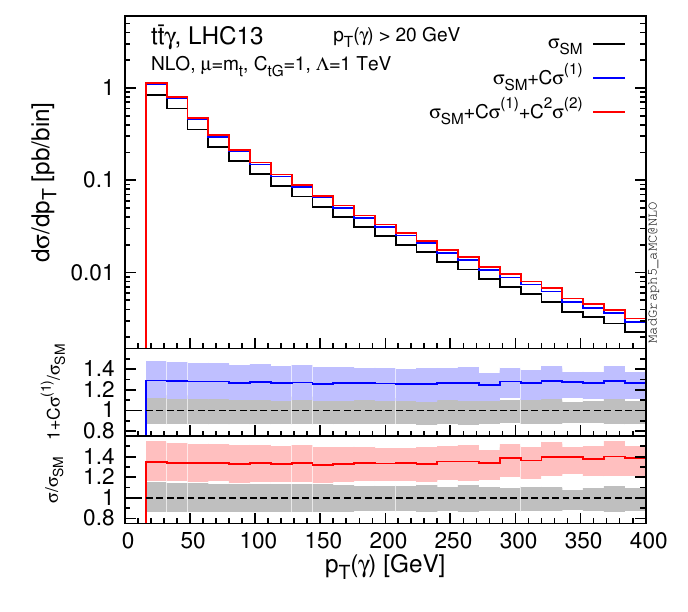}
 \end{minipage}
\caption{\label{fig:ttA_8_13_tg} Invariant mass distributions for the top quark pair and photon $p_T$ distribution at 8 and 13 TeV for the chromomagnetic operator  for $C_{tG}=1$ and $\Lambda=1$~TeV.  Scale uncertainty bands are shown. } 
\end{figure}

For the $\mathcal{O}_{tW}$ and $\mathcal{O}_{tB }$ operators the EFT contributions are very small compared to the SM. In this case we resort to $C_{tB}=4$ to demonstrate the effect of the $\mathcal{O}_{tB }$ operator in Fig.~\ref{fig:ttz_8_13_tB}. For this operator  the interference with the SM  is much smaller than the  $\mathcal{O}(\Lambda^{-4})$ terms which are rising with the energy, as evident in the ratio plots. 
 
\begin{figure}[h]
\centering
\includegraphics[trim=1cm 0 0 0, scale =0.7]{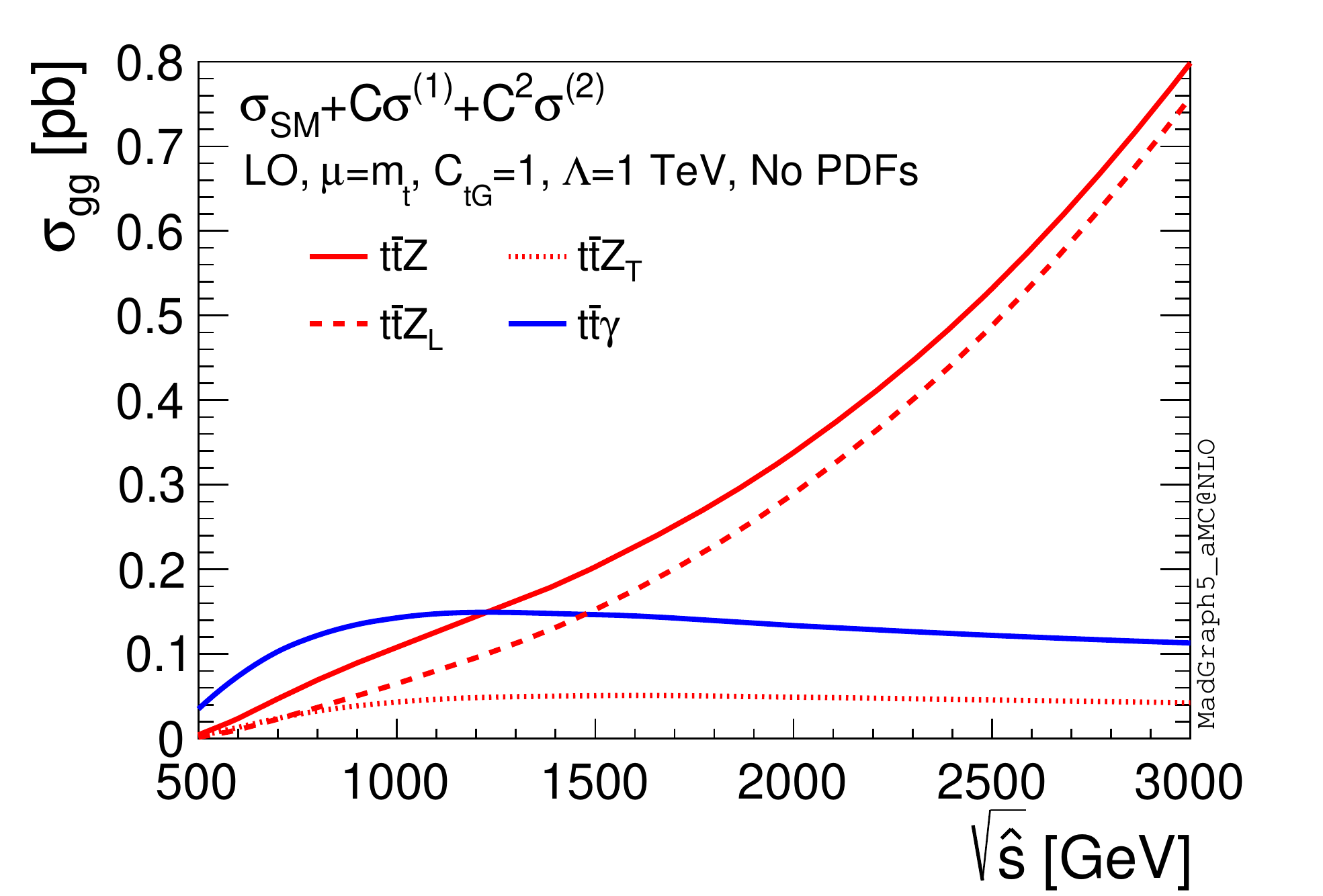}
\label{mttx_tg_8}
\caption{\label{fig:ttA_vs_ttZ} Partonic cross section for $t\bar{t} \gamma$ and $t\bar{t}Z$ as function of the centre of mass energy for the chromomagnetic operator. The $t\bar{t}Z$ cross section is decomposed to transverse and longitudinal $Z$ contributions. } 
\end{figure}

For $t\bar{t}\gamma$, the results for $\mathcal{O}_{tG}$ operators are shown
in Fig.~\ref{fig:ttA_8_13_tg}  for 8 and 13 TeV. We notice that, in contrast with $t\bar{t}Z$, where the squared
terms grow rapidly with the energy, for $t\bar{t}\gamma$ that contribution is
smaller and does not lead to significant changes in the shapes of the two
observables shown here. A comparison of the two processes can be made at the partonic cross-section level as shown in Fig.~\ref{fig:ttA_vs_ttZ}. In this plot the total cross-section is shown, i.e., schematically $\sigma_{SM}+C \sigma^{(1)}+C^2 \sigma^{(2)}$ for the chromomagnetic operator. The  $t\bar{t}Z$ cross-section is decomposed into the transverse and longitudinal $Z$ contributions. 

\noindent
The only component which is rising with the energy is the longitudinal one, which explains why the photon distributions do not show any increase with the energy while those for the $Z$ rise fast. 
In fact in $t\bar{t}Z$, the Higgs field in $\mathcal{O}_{tG}$ always takes its vacuum expectation
value, and so by power counting the squared amplitude scales at most as
$\sim sv^2/\Lambda^4$ for $t\bar{t}Z_T$ and $t\bar{t}\gamma$, which is
not enough for the cross section to rise at high energy.  On the other hand, in
$t\bar{t}Z_L$ the longitudinal polarisation vector contributes an additional
factor of $(E/m_Z)^2$, leading to a final $\sim s^2/\Lambda^4$ scaling of the
squared amplitude.  According to the Goldstone-boson equivalence theorem, the
process $pp\to t\bar{t}G^0$, where $G^0$ is the neutral Goldstone boson, has
the same energy dependence.  This dependence comes solely from the diagram with
a five-point contact interaction, $ggttG^0$, from $\mathcal{O}_{tG}$, and because
here the Higgs field is dynamical by simple power counting the square amplitude
indeed scales as $\sim s^2/\Lambda^4$.  To verify this reasoning, we have
checked that in the process $gt\to tZ_L$, the squared amplitude rises as $\sim
s^2/\Lambda^4$, and the leading term in the energy expansion can be fully
reproduced by a single diagram with a contact $gttG^0$ interaction. An
analogous example of a high-energy growth is discussed in \cite{Dror:2015nkp}
where $tW\to tW$ scattering in $t\bar{t}Wj$ electroweak production is employed
to provide information on the top-$Z$ couplings. 

\begin{figure}[b!]
 \begin{minipage}[t]{0.5\linewidth}
\centering
\includegraphics[trim=1cm 0 0 0, scale =1.2]{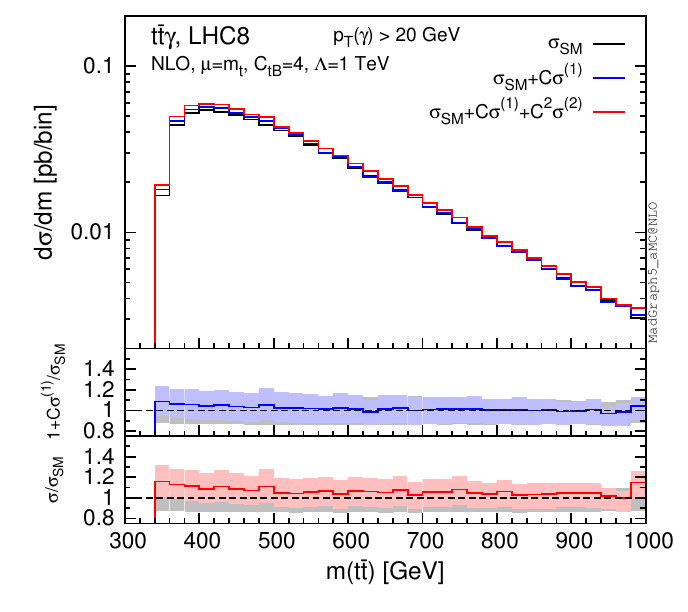}
\label{mttx_tg_8}
\end{minipage}
\hspace{0.5cm}
 \begin{minipage}[t]{0.5\linewidth}
 \centering
 \includegraphics[trim=1cm 0 0 0,scale=1.2]{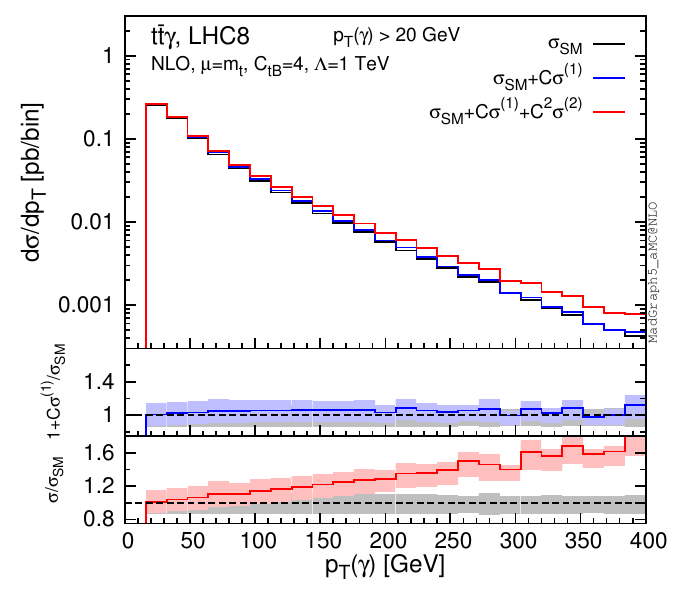}
 \end{minipage}
 \begin{minipage}[t]{0.5\linewidth}
\centering
\includegraphics[trim=1cm 0 0 0, scale =1.2]{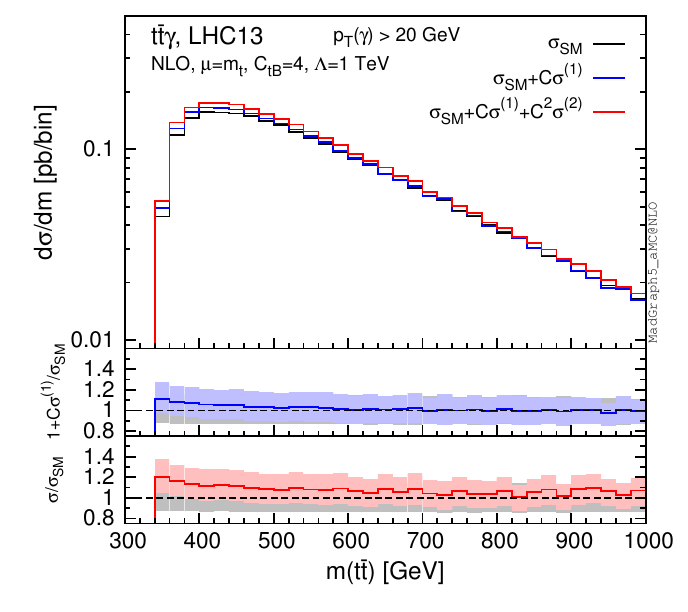}
\label{mttx_tg_8}
\end{minipage}
\hspace{0.5cm}
 \begin{minipage}[t]{0.5\linewidth}
 \centering
 \includegraphics[trim=1cm 0 0 0,scale=1.2]{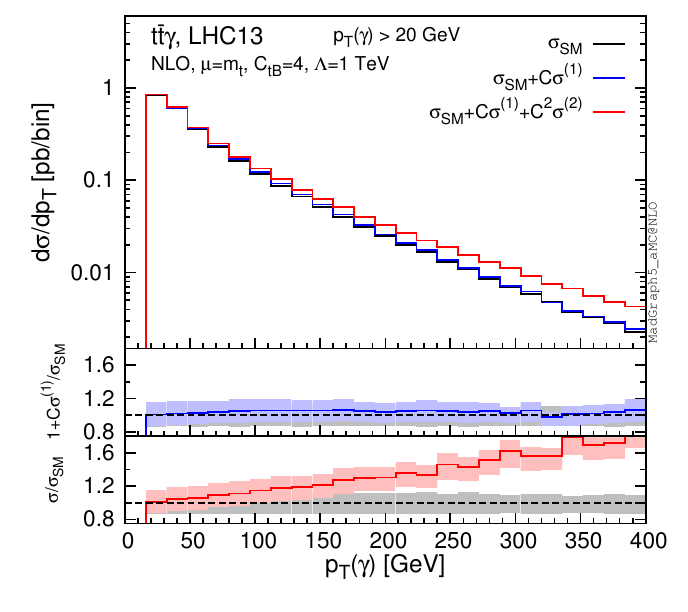}
 \end{minipage}
\caption{\label{fig:ttA_8_13_tB} Invariant mass distributions for the top quark pair and photon $p_T$ distribution at 8 and 13 TeV for the $\mathcal{O}_{tB}$ operator for $C_{tB}=4$ and $\Lambda=1$~TeV.  Scale uncertainty bands are shown. } 
\end{figure}

The corresponding distributions for $\mathcal{O}_{tB}$ are shown in Fig.~\ref{fig:ttA_8_13_tB} for 8 and 13 TeV.  As setting $C_{tB}=1$ does not give any visible deviations from the SM, we resort to $C_{tB}=4$ for these plots.  While the squared term does not rise with $m(t\bar{t})$, it increases fast with  the photon transverse momentum. This is again related to the amount of momentum passing through the EFT vertex. High top pair invariant mass does not correspond to high momentum through the EFT vertex for the $\mathcal{O}_{tB}$ operator, in contrast with the situation for $\mathcal{O}_{tG}$. For $\mathcal{O}_{tG}$ there is a strong correlation between the $m(t\bar{t})$ and the energy in the EFT vertex leading to a rising distribution. 

\begin{figure}[b!]
 \begin{minipage}[t]{0.5\linewidth}
\centering
\includegraphics[trim=1cm 0 0 0, scale =1.2]{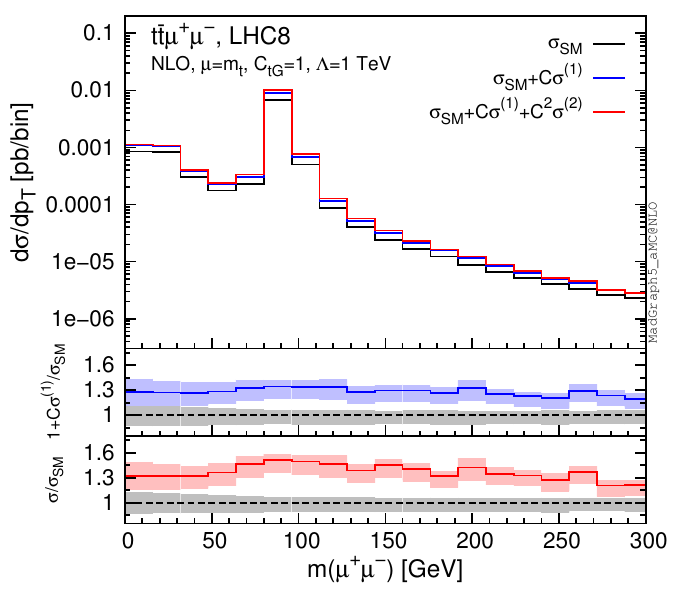}
\label{mttx_tg_8}
\end{minipage}
\hspace{0.5cm}
 \begin{minipage}[t]{0.5\linewidth}
 \centering
 \includegraphics[trim=1cm 0 0 0,scale=1.2]{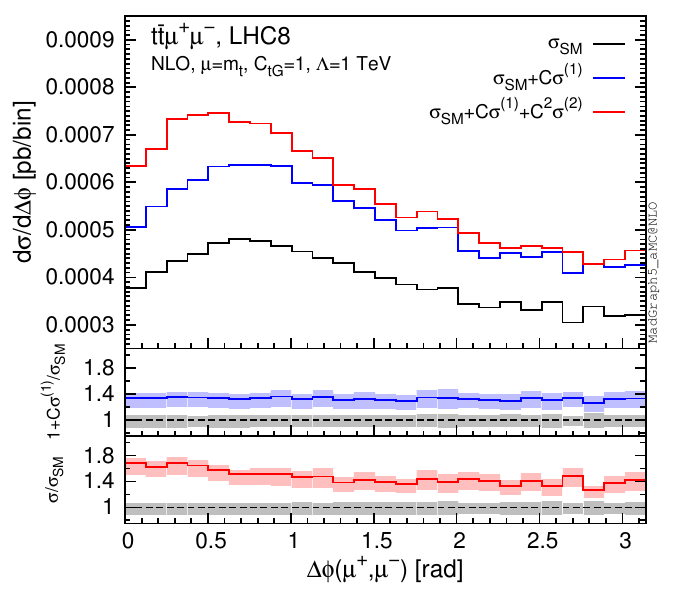}
 \end{minipage}
 \begin{minipage}[t]{0.5\linewidth}
\centering
\includegraphics[trim=1cm 0 0 0, scale =1.2]{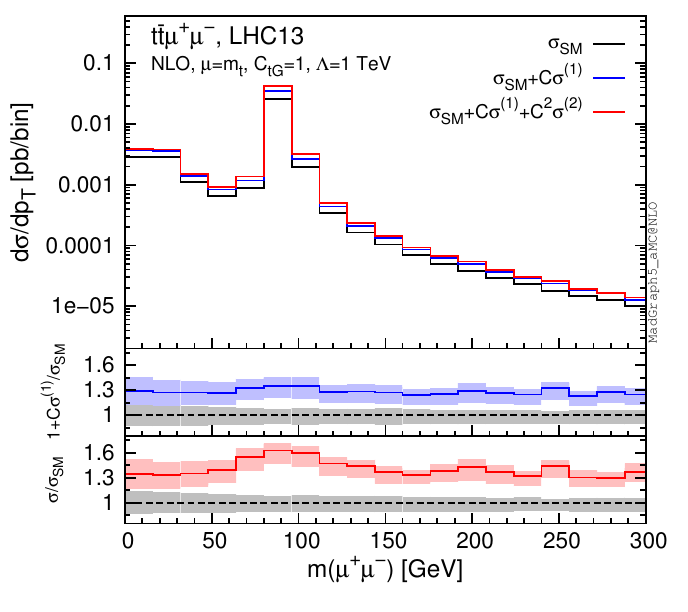}
\label{mttx_tg_8}
\end{minipage}
\hspace{0.5cm}
 \begin{minipage}[t]{0.5\linewidth}
 \centering
 \includegraphics[trim=1cm 0 0 0,scale=1.2]{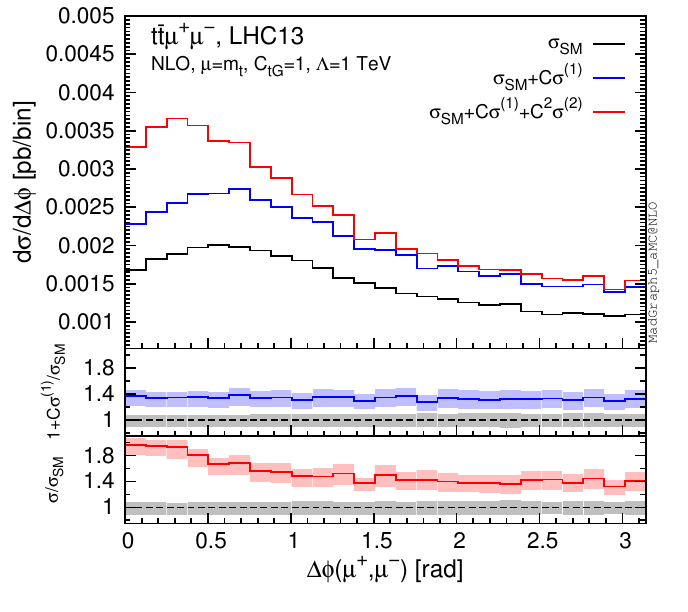}
 \end{minipage}
\caption{\label{fig:ttmunu_8_13_tg} Invariant mass distributions for the lepton pair and lepton angular separation distribution at 8 and 13 TeV for the chromomagnetic operator for the $\mathcal{O}_{tG}$ operator for $C_{tG}=1$ and $\Lambda=1$~TeV. } 
\end{figure}

For the $t\bar{t} \mu^+ \mu^-$ process, we examine the angular separation
between the leptons $\Delta \phi$ and the invariant mass distribution of the two
leptons $m(\ell \ell)$ for the $\mathcal{O}_{tG}$ operator in Fig.~\ref{fig:ttmunu_8_13_tg} for 8 and 13 TeV. The angular separation
between the two leptons is highly correlated with the
transverse momentum of the vector boson. 

\noindent
This implies that at low $\Delta
\phi$, the behaviour matches that of the high vector transverse momentum region, since for a boosted
vector boson, the leptons are collimated. As expected, the behaviour
close to the $Z$ mass peak resembles that of the $t\bar{t}Z$ process, while at
low invariant mass of the lepton pair it approaches that of $t\bar{t}\gamma$.

\begin{figure}[b!]
 \begin{minipage}[t]{0.5\linewidth}
\centering
\includegraphics[trim=1cm 0 0 0, scale =1.2]{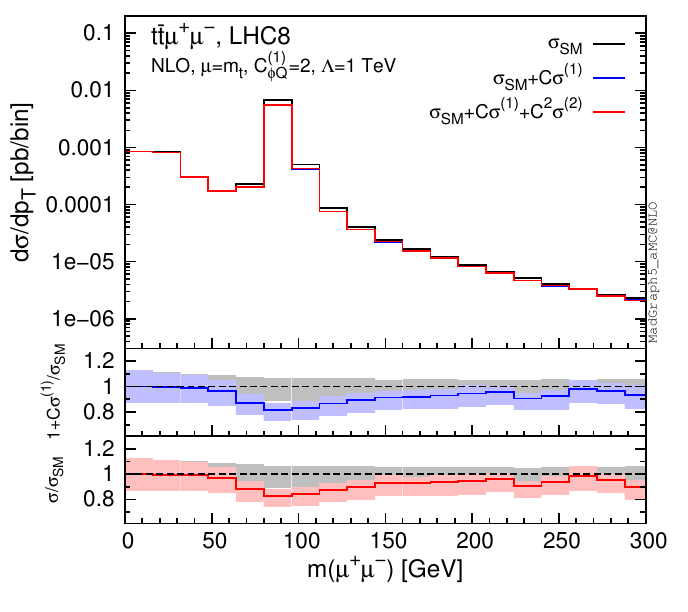}
\label{mttx_tg_8}
\end{minipage}
\hspace{0.5cm}
 \begin{minipage}[t]{0.5\linewidth}
 \centering
 \includegraphics[trim=1cm 0 0 0,scale=1.2]{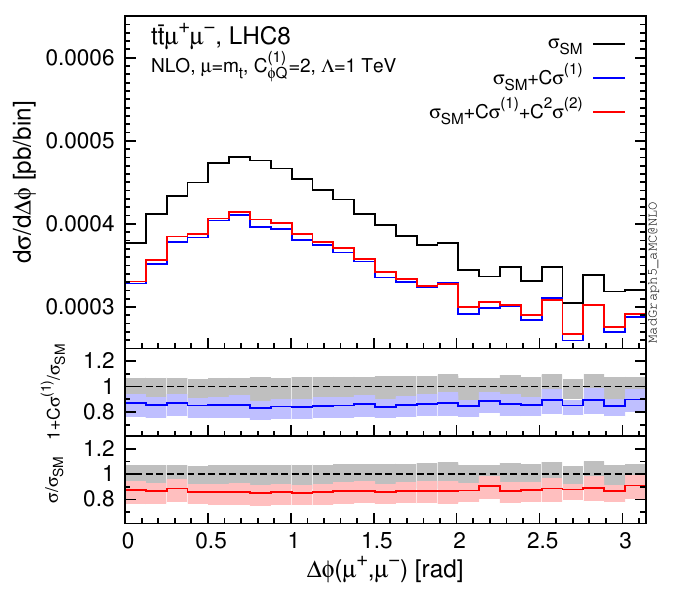}
 \end{minipage}
 \begin{minipage}[t]{0.5\linewidth}
\centering
\includegraphics[trim=1cm 0 0 0, scale =1.2]{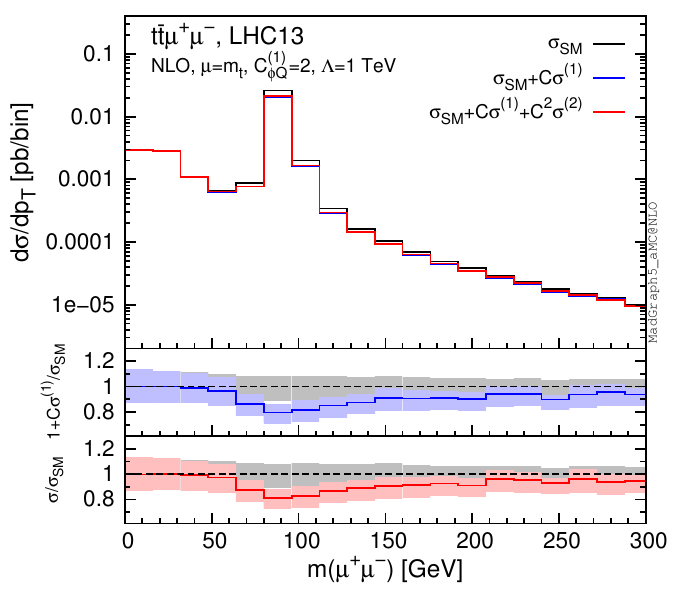}
\label{mttx_tg_8}
\end{minipage}
\hspace{0.5cm}
 \begin{minipage}[t]{0.5\linewidth}
 \centering
 \includegraphics[trim=1cm 0 0 0,scale=1.2]{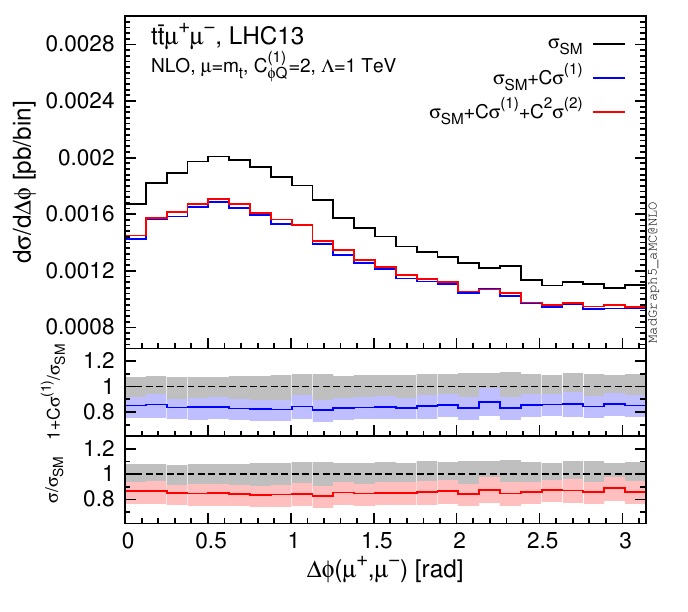}
 \end{minipage}
\caption{\label{fig:ttmunu_8_13_fQ1} Invariant mass distributions for the lepton pair and lepton angular separation distribution at 8 and 13 TeV for the $\mathcal{O}_{\phi Q}^{(1)}$ operator for $C_{\phi Q}^{(1)}=2$ and $\Lambda=1$~TeV.  } 
\end{figure}

The corresponding results for $\mathcal{O}_{\phi Q}^{(1)}$ are shown in Fig.~\ref{fig:ttmunu_8_13_fQ1}. Again the behaviour of the ratios follows that of the $t\bar{t}Z$ close to the $Z$ mass peak, while at low masses the dimension-six contribution approaches zero as this operator has no effect on the $t\bar{t}\gamma^*$ process which dominates at low $m(\ell\ell)$. The $\Delta \phi$ distributions are rather flat similarly to those of the $p_T(Z)$ for the same operator. For brevity we do not show the results for the rest of the current operators, as they are similar to    $\mathcal{O}_{\phi Q}^{(1)}$.

\begin{figure}[h]
 \begin{minipage}[t]{0.5\linewidth}
\centering
\includegraphics[trim=1cm 0 0 0, scale =1.2]{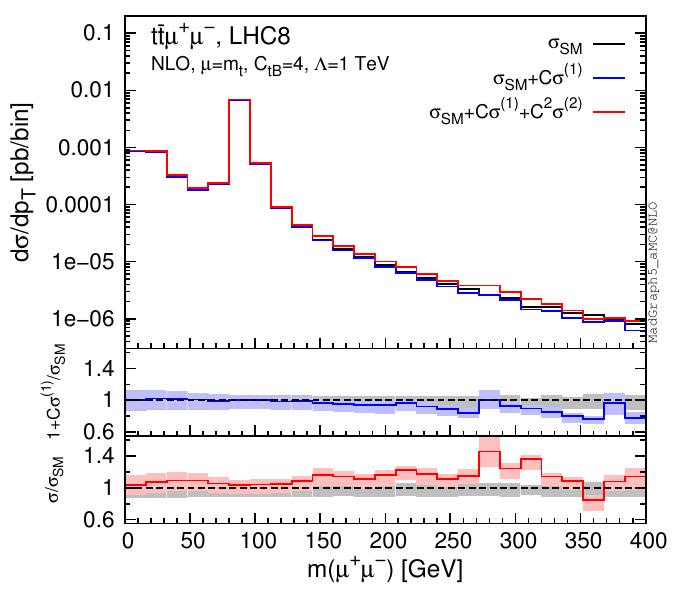}
\label{mttx_tg_8}
\end{minipage}
\hspace{0.5cm}
 \begin{minipage}[t]{0.5\linewidth}
 \centering
 \includegraphics[trim=1cm 0 0 0,scale=1.2]{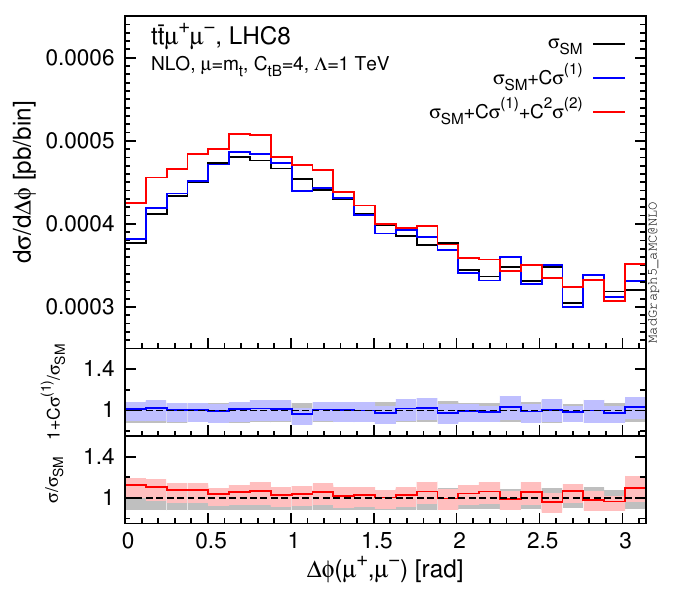}
 \end{minipage}
 \begin{minipage}[t]{0.5\linewidth}
\centering
\includegraphics[trim=1cm 0 0 0, scale =1.2]{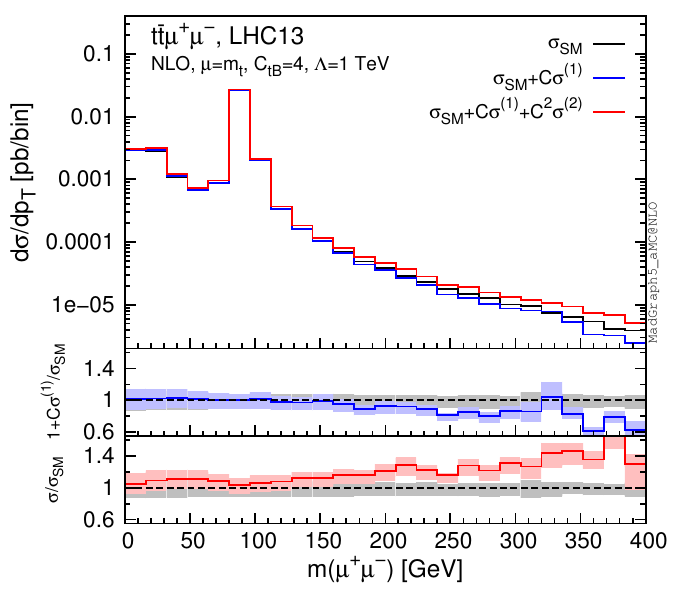}
\label{mttx_tg_8}
\end{minipage}
\hspace{0.5cm}
 \begin{minipage}[t]{0.5\linewidth}
 \centering
 \includegraphics[trim=1cm 0 0 0,scale=1.2]{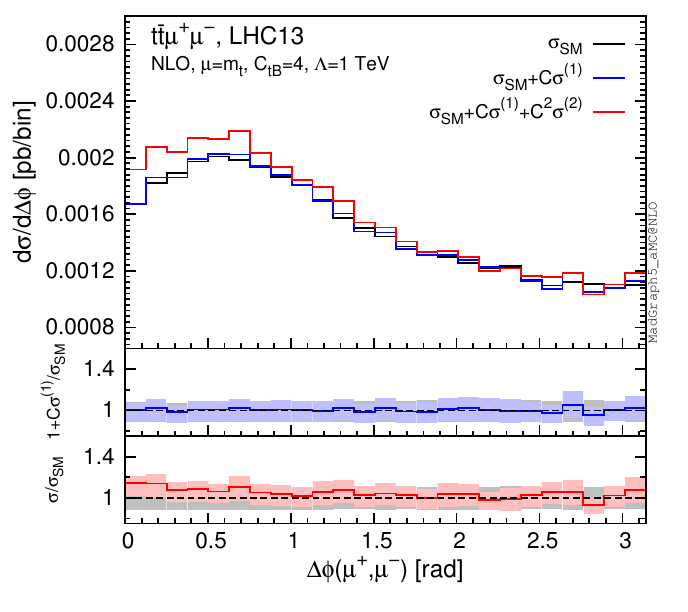}
 \end{minipage}
\caption{\label{fig:ttmunu_8_13_tB} Invariant mass distributions for the lepton pair and lepton angular separation distribution at 8 and 13 TeV for the $\mathcal{O}_{tB}$ operator for $C_{tB}=4$ and $\Lambda=1$~TeV.  } 
\end{figure}

We conclude this section by showing the results for the $\mathcal{O}_{tB}$ operator operator in Fig.~\ref{fig:ttmunu_8_13_tB}. The size of the interference with the SM increases at high lepton pair invariant masses while it is constant as a function of the angular separation between the leptons. The squared terms rise at high invariant mass and low angular separation in agreement with the observations made for the $t\bar{t}\gamma$ and $t\bar{t}Z$ distributions.

\section{Results for $gg\to HZ$}
\begin{figure}[b!]
\centering
\includegraphics[scale=0.7]{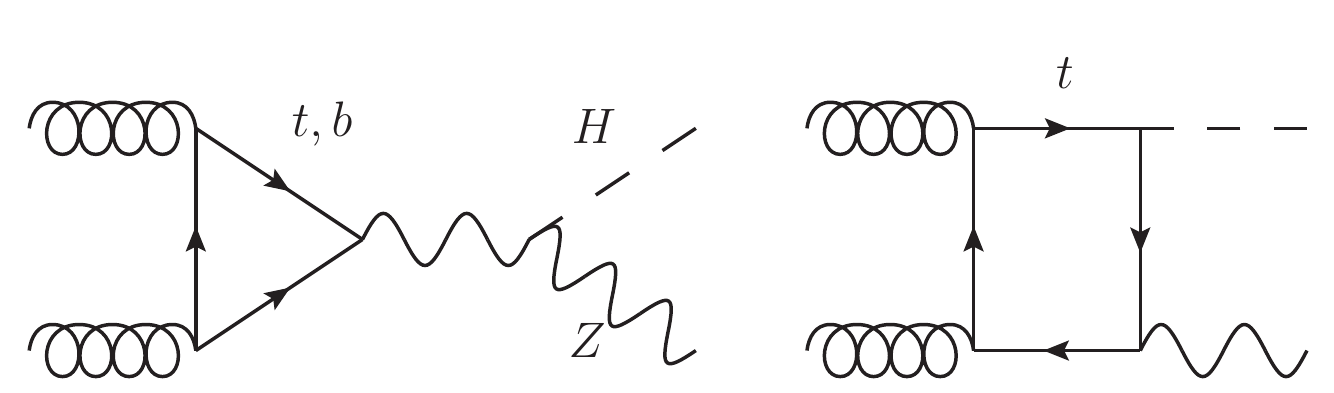}
\caption{Feynman diagrams for $HZ$ production in gluon fusion in the SM. }
\label{fig:ggHZdiagrams}
\end{figure}

A subset of the operators affecting $t\bar{t}Z/t\bar{t}\gamma$ enter also in the associated production of a $HZ$ pair in gluon fusion, shown in the Feynman diagrams of Fig.~\ref{fig:ggHZdiagrams}. This process is formally part of the NNLO cross section for $HZ$ production and contributes at the 10\% level. It is nevertheless particularly important in the high Higgs $p_T$ regions where the experimental searches are most sensitive. This process has been studied within the SM, also including the contribution of additional jet radiation which turns out to be important in the high $p_T$ regions \cite{Hespel:2015zea}. In this work we consider this process as it can provide additional information on the Wilson coefficients once combined with the corresponding $HZ$ measurements at the LHC.  In this section, we investigate the effect of the operators presented above on this process. We note that we 
consider only the operators involving the top quark and ignore all other
dimension-six operators, such as those affecting the interaction of the Higgs with
the vector bosons. 

\begin{figure}[b!]
\centering
\includegraphics[scale=0.7]{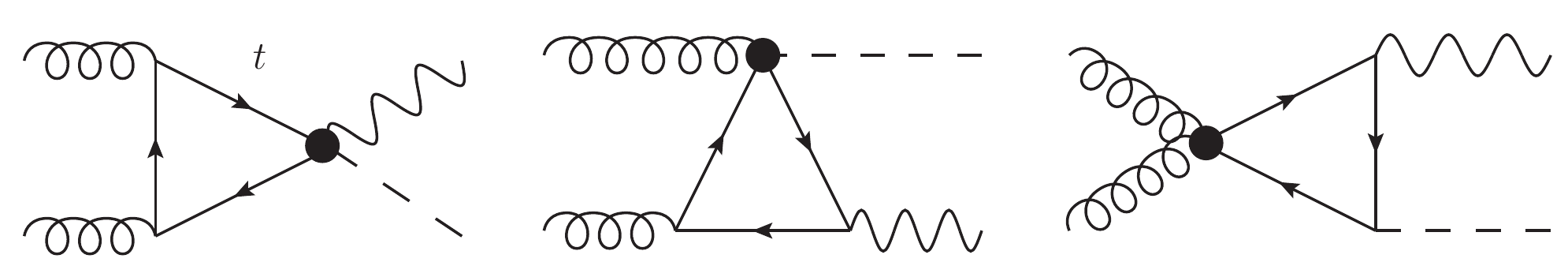}
\caption{Additional types of Feynman diagrams for $HZ$ production in gluon fusion in the presence of dimension 6 operators. The new vertices originating from the dimension-six operators are denoted with a blob. }
\label{fig:ggHZBSM}
\end{figure}

\begin{table}[b!]
\renewcommand{\arraystretch}{1.5}
\small
\begin{center}
\begin{tabular}{c | c | c c c}
\hline
  [fb] & SM &  & $\mathcal{O}_{tG}$ & $\mathcal{O}^{(1)}_{\phi Q}$ \\
 \hline
\multirow{4}{*}{8TeV}  & \multirow{4}{*} {$ 29.15^{ +40.0 \% }_{ -26.6 \% } $} & $\sigma_{i}^{(1)}$ & $ 10.37^{ +41.3 \% }_{ -27.2 \% } $ & $ 1.719^{ +42.5 \% }_{ -27.6 \% } $ \\
 &   & $\sigma_{i}^{(2)}$ & $ 1.621^{ +45.1 \% }_{ -28.7 \% } $ & $ 0.0469^{ +46.5 \% }_{ -29.2 \% } $  \\
  &   & $\sigma_{i}^{(1)}/\sigma_{SM}$ & $ 0.356^{ +0.9 \% }_{ -0.8 \% } $ & $ 0.0590^{ +1.8 \% }_{ -1.4 \% } $ \\
   &   & $\sigma_{i}^{(2)}/\sigma_{i}^{(1)}$ & $ 0.156^{ +2.6 \% }_{ -2.0 \% } $ & $ 0.0273^{ +2.8 \% }_{ -2.3 \% } $  \\
 \hline
\multirow{4}{*}{13TeV}  &  \multirow{4}{*}{$ 93.6^{ +34.3 \% }_{ -23.8 \% } $} & $\sigma_{i}^{(1)}$ &$ 34.6^{ +35.2 \% }_{ -24.5 \% } $ & $ 5.91^{ +36.4 \% }_{ -24.9 \% } $ \\
 &   & $\sigma_{i}^{(2)}$ & $ 6.09^{ +39.2 \% }_{ -26.1 \% } $ & $ 0.182^{ +40.2\% }_{ -26.6 \% } $  \\
  &   & $\sigma_{i}^{(1)}/\sigma_{SM}$ & $ 0.370^{ +0.7 \% }_{ -0.9 \% } $ & $ 0.0631^{ +1.6 \% }_{ -1.5 \% } $ \\
   &   & $\sigma_{i}^{(2)}/\sigma_{i}^{(1)}$ & $ 0.176^{ +2.9 \% }_{ -2.1 \% } $ & $ 0.0309^{ +2.8 \% }_{ -2.2 \% } $ \\
\hline
\end{tabular}
\end{center}
 \caption{\label{tab:gg_HZ} Cross sections (in fb) for $gg \rightarrow HZ$ production at the LHC at $\sqrt{s} =  8$~TeV and $\sqrt{s} =  13$~TeV for the SM and the dimension-six operators. Scale uncertainties are shown in percentages. }  
\end{table}
\noindent
In addition to modifying the interactions in the SM-like diagrams of Fig.~\ref{fig:ggHZdiagrams},
the dimension-six operators introduce additional vertices and hence Feynman
diagrams as shown in Fig.~\ref{fig:ggHZBSM}.  

For this process the factorisation and renormalisation scale is set to
$m_H=125$ GeV. Only LO results can be
obtained as the NLO computation requires 2-loop multi-scale Feynman integrals which are currently not available. The results are shown in Table
\ref{tab:gg_HZ} for both the SM and the dimension-six operators cross sections, the corresponding scale uncertainties and the
corresponding cross-section ratios for 8 and 13 TeV.

\begin{figure}[b!]
 \begin{minipage}[t]{0.5\linewidth}
\centering
\includegraphics[trim=1cm 0 0 0, scale =1.2]{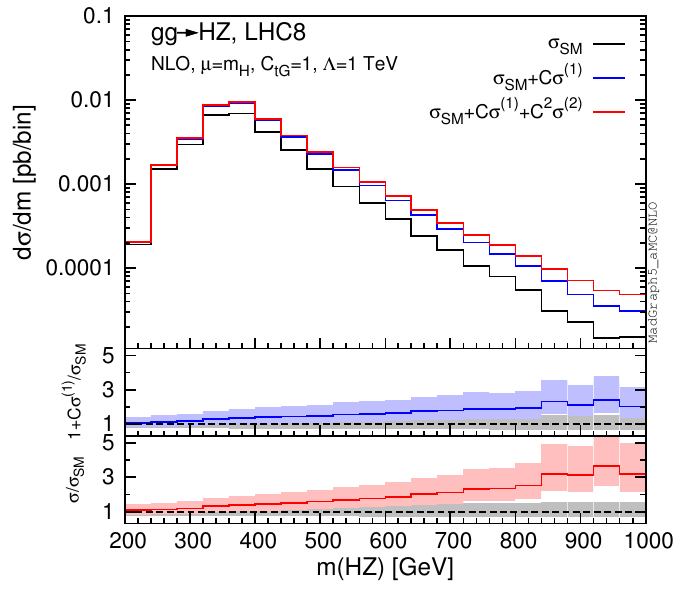}
\end{minipage}
\hspace{0.5cm}
 \begin{minipage}[t]{0.5\linewidth}
 \centering
 \includegraphics[trim=1cm 0 0 0,scale=1.2]{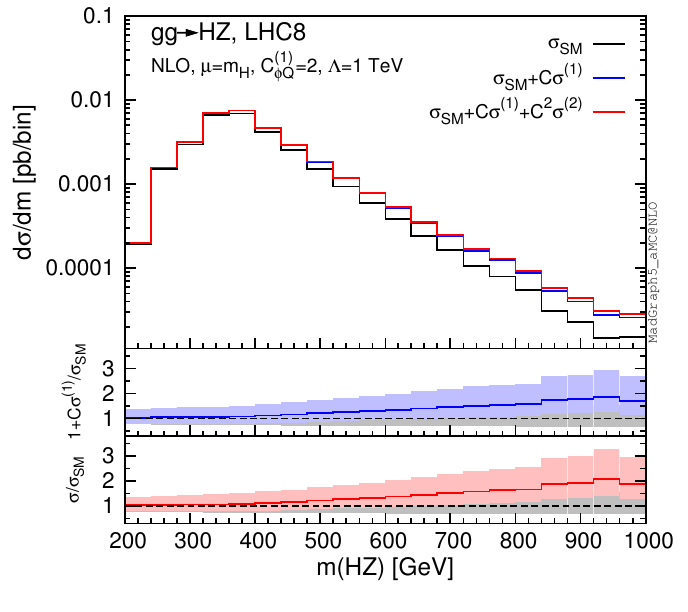}
 \end{minipage}
 \begin{minipage}[t]{0.5\linewidth}
\centering
\includegraphics[trim=1cm 0 0 0, scale =1.2]{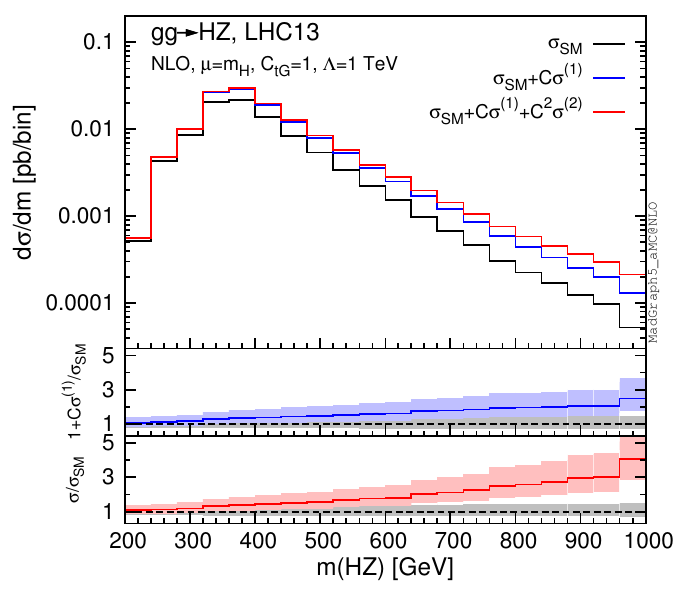}
\end{minipage}
\hspace{0.5cm}
 \begin{minipage}[t]{0.5\linewidth}
 \centering
 \includegraphics[trim=1cm 0 0 0,scale=1.2]{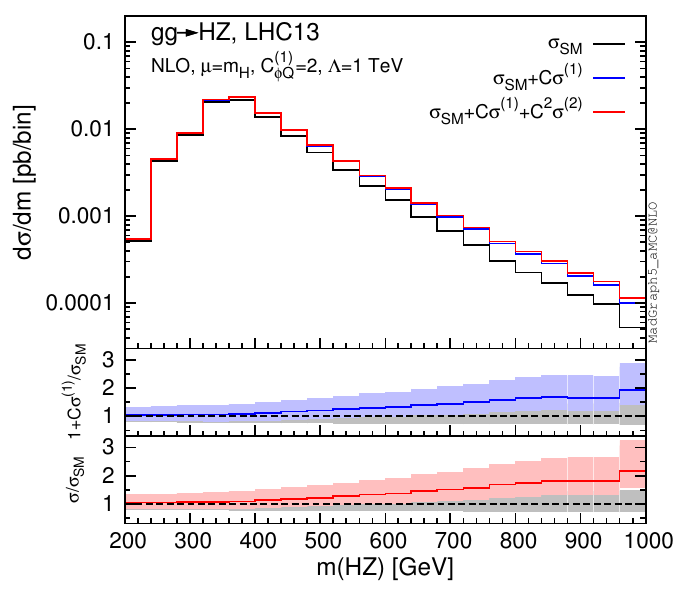}
 \end{minipage}
\caption{\label{fig:ggHZ_8_13} $HZ$ invariant mass distributions for $gg\to HZ$ at 8 and 13 TeV for the $\mathcal{O}_{tG}$ and  $\mathcal{O}_{\phi Q}^{(1)}$ operators. Scale uncertainty bands are shown.} 
\end{figure}
\noindent
The $\mathcal{O}_{tW}$ and $\mathcal{O}_{tB}$ operators do not contribute to
this process, due to charge conjugation invariance. The
$\mathcal{O}^{(3)}_{\phi Q}$, $\mathcal{O}^{(1)}_{\phi Q}$ and
$\mathcal{O}_{\phi t}$ give the same contributions (with a relative minus sign as determined by Eq.~\ref{eq:axialZcoup}) in the
massless $b$-quark limit, as they affect in the same way the axial vector
coupling of the top to the $Z$, which is the only component whose contribution
is allowed by charge conjugation symmetry. If one wants
to cancel the chiral anomaly in the triangle loop diagrams with the $Z$-boson
in the $s$-channel, the $\mathcal{O}_{\phi b}$ operator can be included with its Wilson coefficient  set  to $C_{\phi b}=2C_{\phi Q}^{(1)}-C_{\phi t}$.
By appropriately fixing the coefficient of $\mathcal{O}_{\phi b}$ the axial-vector coupling of the bottom remains opposite to that of the top and the anomaly cancels. In practice this has a negligible numerical effect on the results. The chromomagnetic operator gives a significant contribution reaching 35\% of the SM cross section for $C_{tG}=1$ and $\Lambda=1$ TeV. The three current operators give contributions at the 6\% level. In both cases the contribution of the squared amplitudes are subdominant at the total cross section level. These results suffer from large scale uncertainties as it is often the case with gluon fusion processes at LO. The invariant mass distribution for the $HZ$
pair is shown in Fig.~\ref{fig:ggHZ_8_13} for the SM and the dimension-six operators. For this process we find that both the interference with the SM amplitude and the squared contribution are growing with energy. 

\section{Results for the ILC}
\label{sec:ILC}
The top-quark electroweak couplings can be accurately determined by future $e^+e^-$
colliders, using top-pair production, thanks to the clean background.  
Our approach can be applied to $e^+e^-$ colliders as well, providing more
accurate predictions for deviations that will be measured in this process. 
In this section we present results obtained for the ILC at $\sqrt{s}=500$ GeV
for top pair production. For this process, the $\mathcal{O}_{tG}$ operator
contributes only at NLO, while the other operators contribute starting at LO.
The results are presented in Table \ref{tab:sigmattILC}. In this case, we do
not show the renormalisation scale uncertainties as these can be computed only
at NLO and are at the 1-2\% level. 

\begin{table}[h]
\begin{center}
\begin{tabular}{ccccccccc}
\hline
   500GeV & SM &  $\mathcal{O}_{tG}$ & $\mathcal{O}^{(3)}_{\phi Q}$ &  $\mathcal{O}^{(1)}_{\phi Q}$ &  $\mathcal{O}_{\phi t}$ &  $\mathcal{O}_{tW}$ &  $\mathcal{O}_{tB}$ \\
\hline
 $\sigma_{i,LO}^{(1)}$  & 566  & 0 & 15.3 & -15.3  & -1.3 & 272  & 191 \\
  $\sigma_{i,NLO}^{(1)}$ &  647 & -6.22 & 18.0 & -18.0 & -1.0 & 307 & 216  \\
 K-factor  &1.14 & N/A  &1.17  & 1.17   & 0.78 & 1.13 & 1.13 \\
  $\sigma_{i,LO}^{(2)}$  &   & 0 & 0.72 & 0.71  & 0.72 & 60.4  & 27.2 \\
  $\sigma_{i,NLO}^{(2)}$ &   & 0.037  & 0.83 & 0.82 & 0.82 & 68.8 & 31.0  \\
 $\sigma_{i,LO}^{(1)}/\sigma_{SM,LO}$ & & 0 & 0.027  & -0.027   & -0.0022  & 0.48  & 0.34 \\
 $\sigma_{i,NLO}^{(1)}/\sigma_{SM,NLO}$  & & -0.096  &  0.028 &-0.028 & -0.0015 & 0.47 & 0.33 \\
 $\sigma_{i,LO}^{(2)}/\sigma_{i,LO}^{(1)}$ & & N/A & 0.047 & -0.047 & -0.57 & 0.22 & 0.14 \\
 $\sigma_{i,NLO}^{(2)}/\sigma_{i,NLO}^{(1)}$ & & -0.006 & 0.046 & -0.046 & -0.82 & 0.22 & 0.14  \\
\hline
\end{tabular}
\end{center}
 \caption{\label{tab:sigmattILC} Cross sections (in fb) for $t\bar{t}$ production at the ILC at $\sqrt{s} =  500$~GeV. Renormalisation scale uncertainties are not shown. They are only present at NLO and remain at the 1\% level. }  
\end{table}

Unlike the $t\bar tV$ processes, here we find significant contributions from the
dipole operators $\mathcal{O}_{tB}$ and $\mathcal{O}_{tW}$, while the other
operators are suppressed, with $\mathcal{O}_{tG}$, $\mathcal{O}_{\phi Q}^{(1)}$
and $\mathcal{O}_{\phi Q}^{(3)}$ at the percent level, and $\mathcal{O}_{\phi
t}$ at the per mille level.  This is
mainly because the momenta of $Z$ and $\gamma$ are at least at the $t\bar t$
threshold, and so the same dipole structure, which suppresses $t\bar tV$ production
at the LHC, enhances the $t\bar t$ production at the ILC.  It follows that
the ILC could provide useful information complementary to the LHC as discussed
also in \cite{Rontsch:2014cca,Rontsch:2015una}. We note here that the analysis of \cite{Rontsch:2014cca,Rontsch:2015una} does not include the contribution from $\mathcal{O}_{tG}$, although (following an anomalous coupling approach) it does include the contribution of the squares of the amplitudes with the top anomalous couplings and therefore also the CP-odd contributions. 

\section{Theoretical uncertainties}
\label{sec:unc}
In this section we briefly discus various theoretical uncertainties relevant to
our results.  In the SMEFT calculation there are two main types of theoretical
uncertainties, those related to missing higher orders in the strong coupling and 
those from higher terms in the $1/\Lambda$ expansion. In the former class, we can list

\begin{itemize}

	\item Uncertainties due to parton-distribution functions. 
		
	This type of uncertainty is also present in the SM calculations and can be treated in the same way, i.e. by following the procedures associated with 
	the corresponding PDF sets, as long as
	the scale of new physics is high enough and the EFT operators do not modify the DGLAP equations. 

	\item Uncertainties due to missing higher orders in the $\alpha_s$
		expansion as in the SM. 
		
	This kind of uncertainty is typically estimated by varying the
	renormalisation and factorisation scales as done in SM calculations.  
	All results presented in this work are provided along with uncertainties that
	are estimated by varying these two scales independently.
	
	\item Uncertainties due to missing higher orders in the $\alpha_s$
		expansion of the EFT operators. 
		
	In the SMEFT an additional uncertainty, related to the scale at which
	the operators are defined, should be considered as well. It
	characterises the uncancelled logarithmic terms in the renormalisation
	group running and mixing of the operators. We did not evaluate these
	uncertainties explicitly even though it is possible in our framework.
	For the operators we have studied in this work, they are expected to be
	negligible compared to the first two scale
	uncertainties~\cite{Zhang:2016omx}.  This is because the anomalous
	dimensions of the relevant operators happen to be smaller by roughly an
	order of magnitude compared to the beta function of  $\alpha_s$
	(see Ref.~\cite{Zhang:2016omx} for a discussion of the operator scale
	uncertainty in the single-top processes).

\end{itemize}
We now consider uncertainties due to missing $\mathcal{O}(\Lambda^{-4})$ contributions.  
Up to this order, the cross section (or any other observable) can be written as:
	\begin{flalign} 
		\sigma=\sigma_{SM}+\sum_i
		\frac{C_i^\mathrm{dim6}}{(\Lambda/1\mathrm{TeV})^2}\sigma_i^{(1,\mathrm{dim6})}
		+\sum_{i<j}
		\frac{C_i^\mathrm{dim6}C_j^\mathrm{dim6}}{(\Lambda/1\mathrm{TeV})^4}\sigma_{ij}^{(2,\mathrm{dim6})}
		+\sum_i
		\frac{C_i^\mathrm{dim8}}{(\Lambda/1\mathrm{TeV})^4}\sigma_i^{(1,\mathrm{dim8})}
	\end{flalign}
The last two terms are formally $\mathcal{O}(\Lambda^{-4})$ contributions, and
may be neglected as they are expected to be suppressed for $\mathcal{O}(1)$ coefficients. 
One should then consider
	
\begin{itemize}

	\item Impact of the squared contributions $\sigma_{ij}^{(2,\mathrm{dim6})}$ coming from dimension-six operators.
	
	These contributions can be explicitly calculated with our approach,
	even though obtaining the complete results can be time consuming.
	In this work we have always provided the results for
	$\sigma^{(2)}_{ii}$ for each operator $\mathcal{O}_i$, for not only
	total cross sections but also for distributions.  In fact, one could
	include these squared contributions in the central values as part of the 
	theoretical predictions, if only one operator is taken to be non-zero
	at a time.  As we have mentioned, this can be justified for cases where
	the expansion in
	$E^2/\Lambda^2$ is under control but the squared contribution may still
	be large, due to less constrained operator coefficients, i.e.~if $C_i^2
	\frac{E^4}{\Lambda^4}>C_i\frac{E^2}{\Lambda^2}>1>\frac{E^2}{\Lambda^2}$
	is satisfied.
	In any case, our results for  the $\sigma^{(2)}_{ii}$ terms can provide
	useful information for the evaluation of the uncertainties, if the
	squared contributions are neglected or only partly included.

	As we have discussed already, the relative size of $\sigma^{(2)}_{ii}$
	compared to $\sigma^{(1)}_{i}$ does not imply anything about the
	validity of the EFT and careful assessment should be done on a
	case-by-case basis.
	
	\item Validity of the EFT, i.e.~contributions from missing higher-dimensional operators.

	The second contribution at $\mathcal{O}(\Lambda^{-4})$,
	$\sigma_{i}^{(1,\mathrm{dim8})}$, comes from interference between SM and
	dimension-eight operators.  These contributions cannot be computed in
	our approach, and will have to be neglected.  A corresponding
	uncertainty should be taken into account.  This can be easily done at
	the LO by calculating the interference contribution from typical
	dimension-eight operators.  Alternatively, by simple power counting,
	these uncertainties may be estimated to be of order
	$C^{\mathrm{dim6}}_i/(\Lambda/1\mathrm{TeV})^2
	\sigma_i^{(1,\mathrm{dim6})} s/\Lambda^2$.  In this work we do not
	assume a specific value of $\Lambda$, and so evaluating such uncertainties is
	not possible without additional assumptions.  However, in a real
	analysis, for any given $\Lambda$, one can always apply a cut
	$s_\mathrm{max}$ on the centre-of-mass energy of the process, so that
	this uncertainty remains under control. 
\end{itemize}

\section{Discussion}
\label{sec:disc}
In this section we explore the sensitivity of the top processes discussed above
to the various operators.  Experimental results from
\cite{CMS:2014gta,Khachatryan:2015sha,Aad:2015eua,CMS:2014wma} are used.
For \cite{Aad:2015eua} and \cite{CMS:2014wma} a direct comparison is
difficult, because of the way in which the measured cross sections are
defined.  We thus define the ``$R$'' ratios in order to facilitate a direct
comparison between the quoted experimental measurements and our theory predictions, as explained in appendix~B.  These ratios are always taken into
account when experimental results on $t\bar t\mu^+\mu^-$ and $t\bar t\gamma$
are used.  On the other hand, the other measurements can be directly compared
with our predictions.

\begin{figure}[h]
\centering
\includegraphics[scale=0.7]{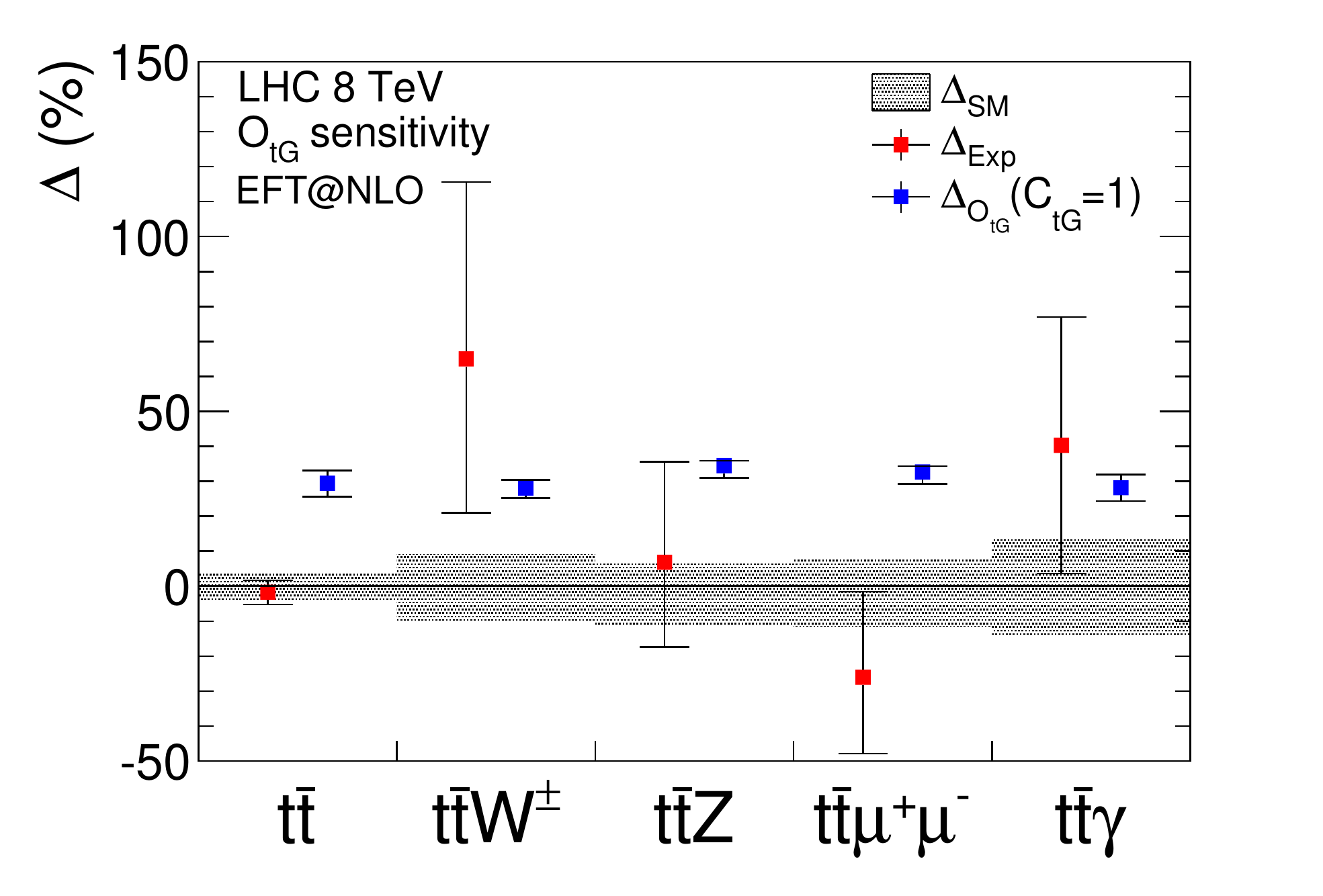}
\caption{Sensitivity of various processes to the $\mathcal{O}_{tG}$ operator.
$\Delta$ denotes the percentage difference from the SM theoretical prediction
for each process. Theory predictions for all $t\bar{t}V$ processes are at NLO
in QCD while for $t\bar{t}$ the NNLO result of \cite{Czakon:2013goa} is
employed. Experimental measurements are also shown along with the corresponding
experimental uncertainties taken from \cite{CMS:2014gta} for $t\bar{t}$,
\cite{Khachatryan:2015sha} for $t\bar{t}W$ and $t\bar{t}Z$, \cite{Aad:2015eua}
for $t\bar{t}\mu^+\mu^-$ and \cite{CMS:2014wma} for $t\bar{t}\gamma$. }
\label{fig:ctgsens}
\end{figure}

We first examine the $\mathcal{O}_{tG}$ operator, which affects all $t\bar{t}V$
processes as well as top pair production. The sensitivity of various processes
to the $\mathcal{O}_{tG}$ operator is demonstrated in Fig.~\ref{fig:ctgsens}.
In the plot we include the percentage deviation from the SM predictions for top
pair production, and top pair production in association with a $W,Z$ boson or a
photon, as well as the $t\bar{t}\ell^+ \ell^-$ process for $C_{tG}=1$ and
$\Lambda =1$ TeV. All SM predictions and uncertainties are NLO, apart from the
top pair production cross-section which is given at NNLO+NNLL
\cite{Czakon:2013goa}. We present also the experimental measurements and the
corresponding uncertainties (systematic and statistical uncertainties added
in quadrature).  Only the $\mathcal{O}(1/\Lambda^2)$
contribution is included. The $\mathcal{O}_{tG}$ operator affects in a similar way all
processes considered here, at the 30\% level for $\Lambda=1$ TeV and
$C_{tG}=1$. At present the most stringent direct constraints on this operator
are obtained from the top pair production measurement which is by far the most
accurate one.

\begin{figure}[t!]
\centering
\includegraphics[scale=0.625]{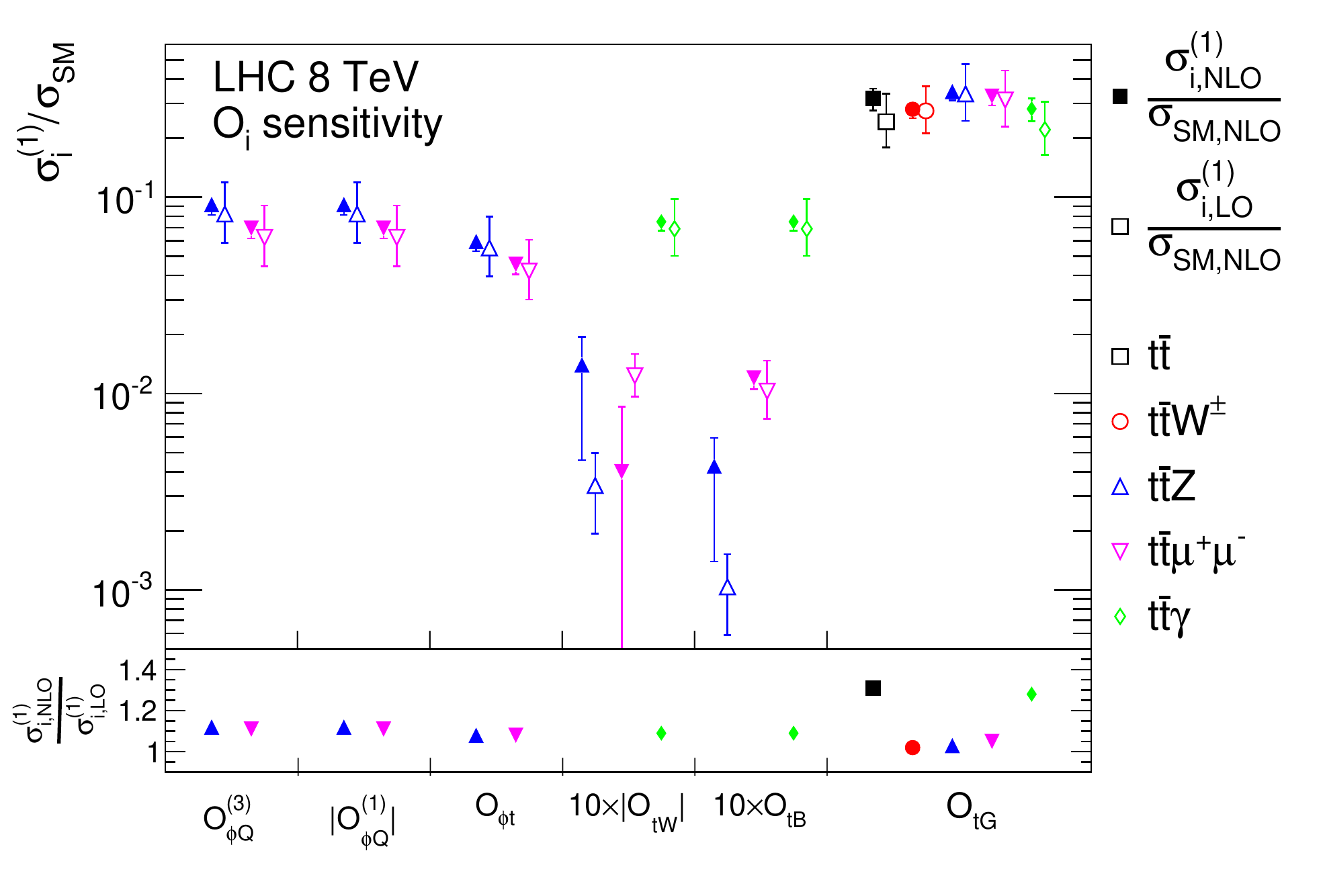}
\caption{Sensitivity of various top quark processes to the various operators shown at LO and NLO at 8 TeV. K-factors are also shown for $\sigma^{(1)}_i$ as well as the scale uncertainties. We do not show the K-factors for the $\mathcal{O}_{tB}$ and $\mathcal{O}_{tW}$ operators in the  $t\bar{t}Z$ and $t\bar{t}\mu^+\mu^-$ processes, as in this case accidental cancellations lead to large or even negative K-factors. }
\label{fig:sensall8}
\end{figure}

\begin{figure}[b!]
\centering
\includegraphics[scale=0.625]{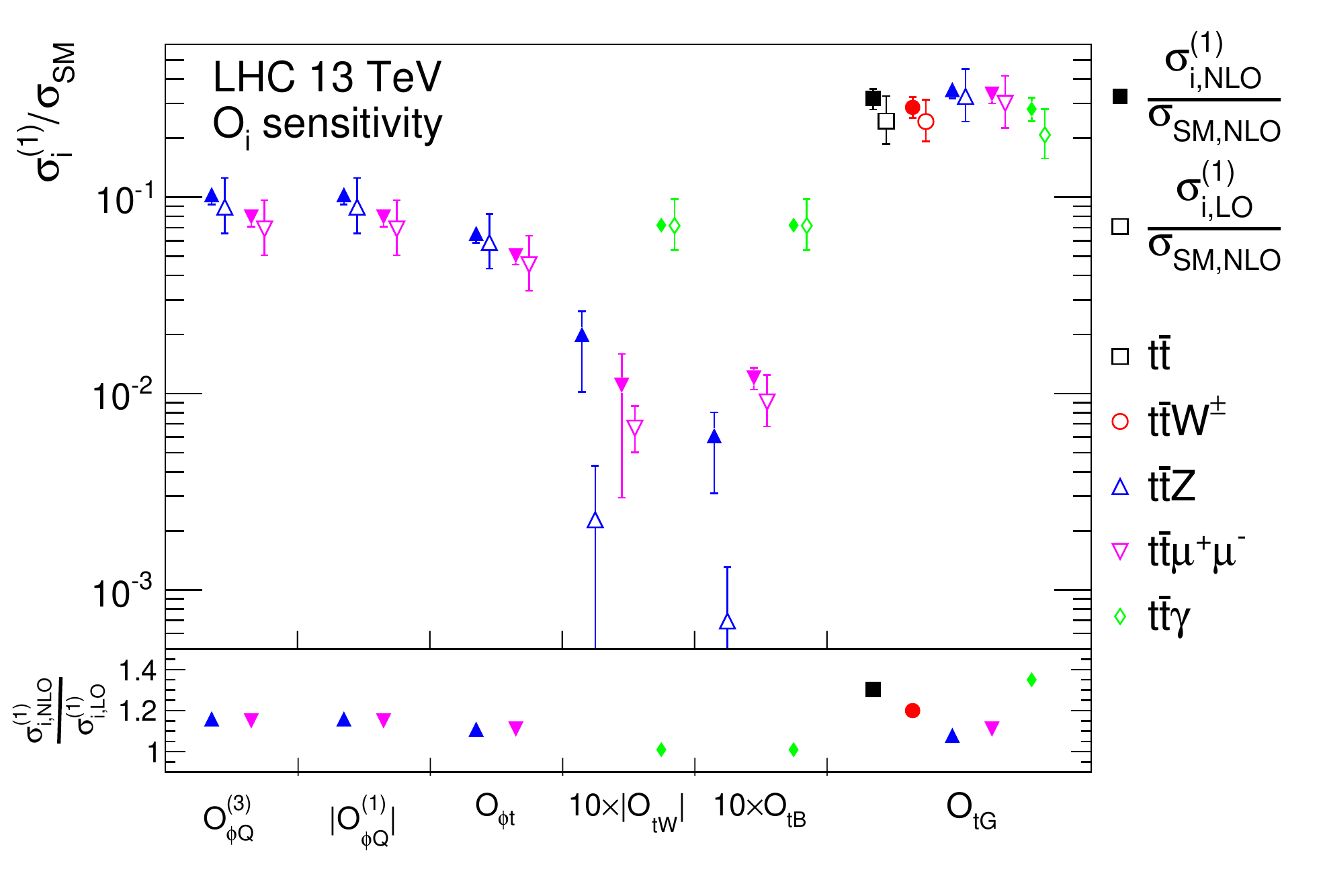}
\caption{Sensitivity of various top quark processes to the various operators shown at LO and NLO at 13 TeV. Details as in Fig.~\ref{fig:sensall8}. }
\label{fig:sensall13}
\end{figure}

The relative sensitivity of the top processes to all operators can be
summarised in Fig.~\ref{fig:sensall8}, where the results for
$C=1$ are shown as a ratio over the SM NLO cross sections,
for the 6 operators considered here both at LO and NLO, along with the
corresponding K-factors in the lower panel. The reduction of the theoretical uncertainties at NLO is also evident in the plot. The corresponding sensitivity plot for 13 TeV is shown in Fig.~\ref{fig:sensall13}, in which similar observations can be made.

\begin{figure}[h]
\centering
\includegraphics[scale=0.74]{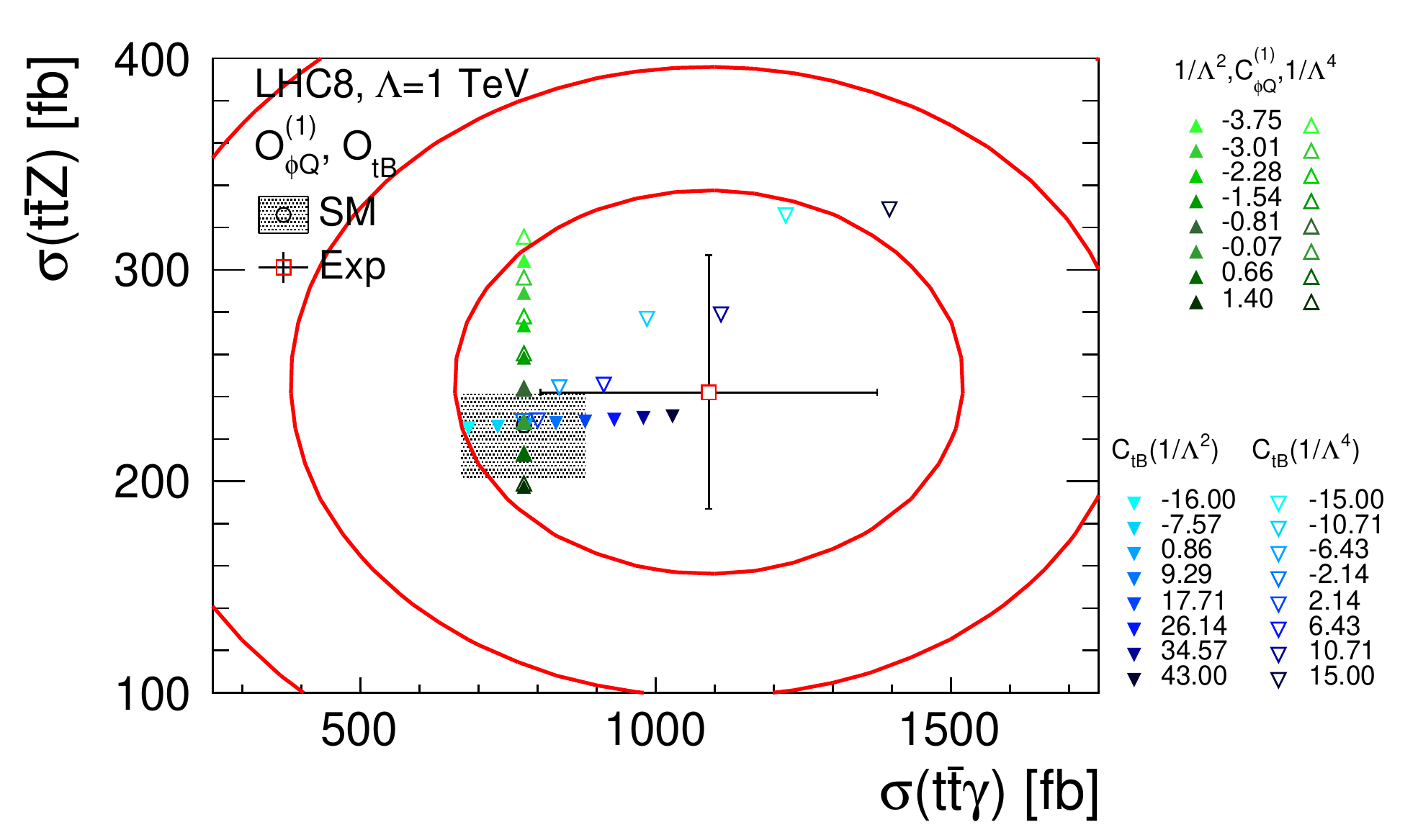}
\caption{Sensitivity of the $t\bar{t}\gamma$ and $t\bar{t}Z$ processes to the $\mathcal{O}^{(1)}_{\phi Q}$ and $\mathcal{O}_{tB}$ operators. For each value of the coefficient we show the cross-section including i) only the interference term (filled triangles) and ii) both the interference and the squared contribution (unfilled triangles). The range for the Wilson coefficients is determined by the current constraints as discussed in Section 2. The experimental measurements used in this plot are taken from \cite{CMS:2014wma} and \cite{Khachatryan:2015sha} for $t\bar{t}\gamma$ and $t\bar{t}Z$ respectively. The squared contribution of the $\mathcal{O}_{tB}$ operator is very large, and therefore we employ a separate smaller interval to obtain cross sections within the boundaries of this plot. }
\label{fig:fq1tB}
\end{figure}

\begin{figure}[h]
\centering
\includegraphics[scale=0.685]{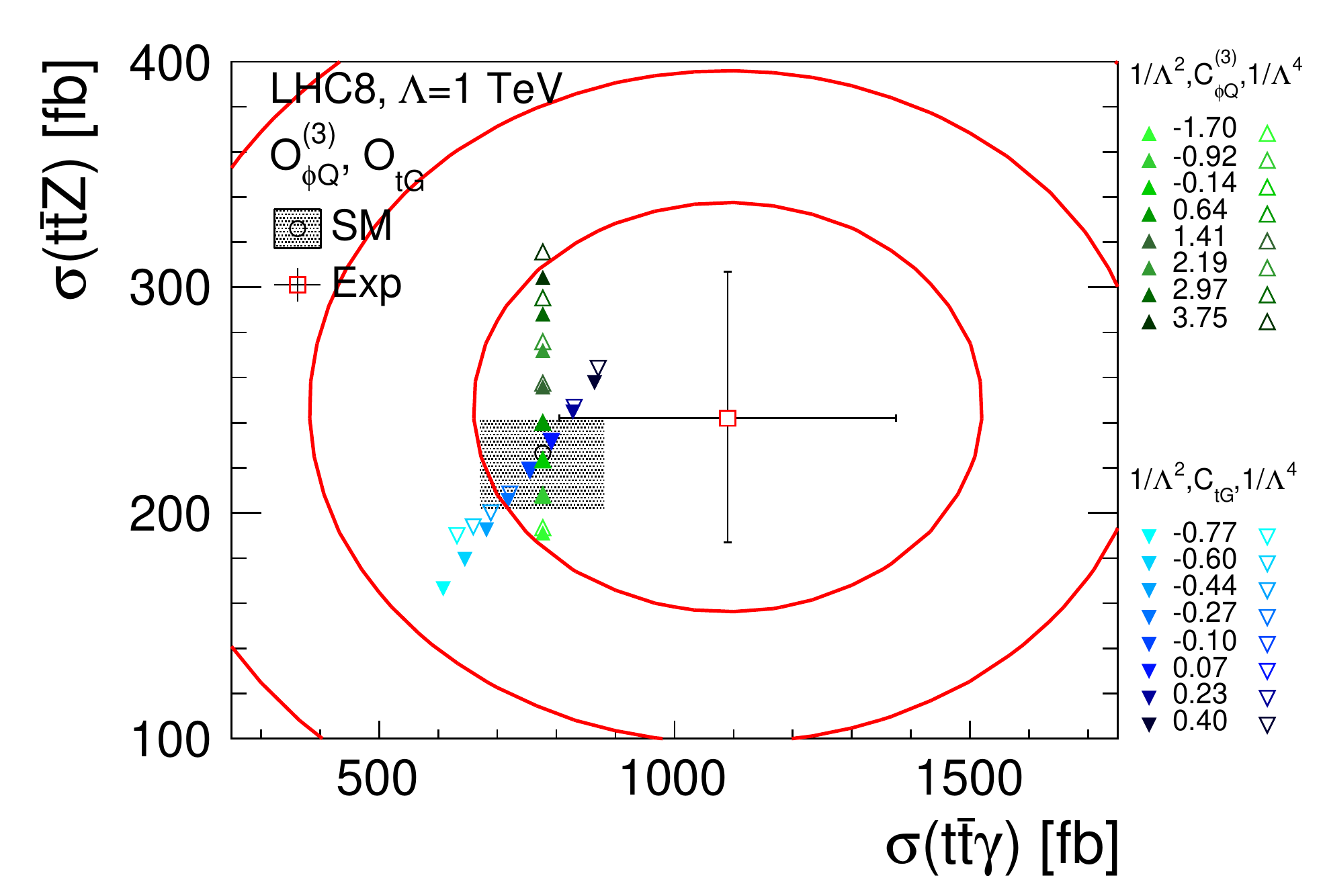}
\caption{Sensitivity of the $t\bar{t}\gamma$ and $t\bar{t}Z$ processes to the $\mathcal{O}^{(3)}_{\phi Q}$ and $\mathcal{O}_{tG}$ operators. Details as in Fig.~\ref{fig:fq1tB}. }
\label{fig:tgfq3}
\end{figure}

\begin{figure}[h]
\centering
\includegraphics[scale=0.685]{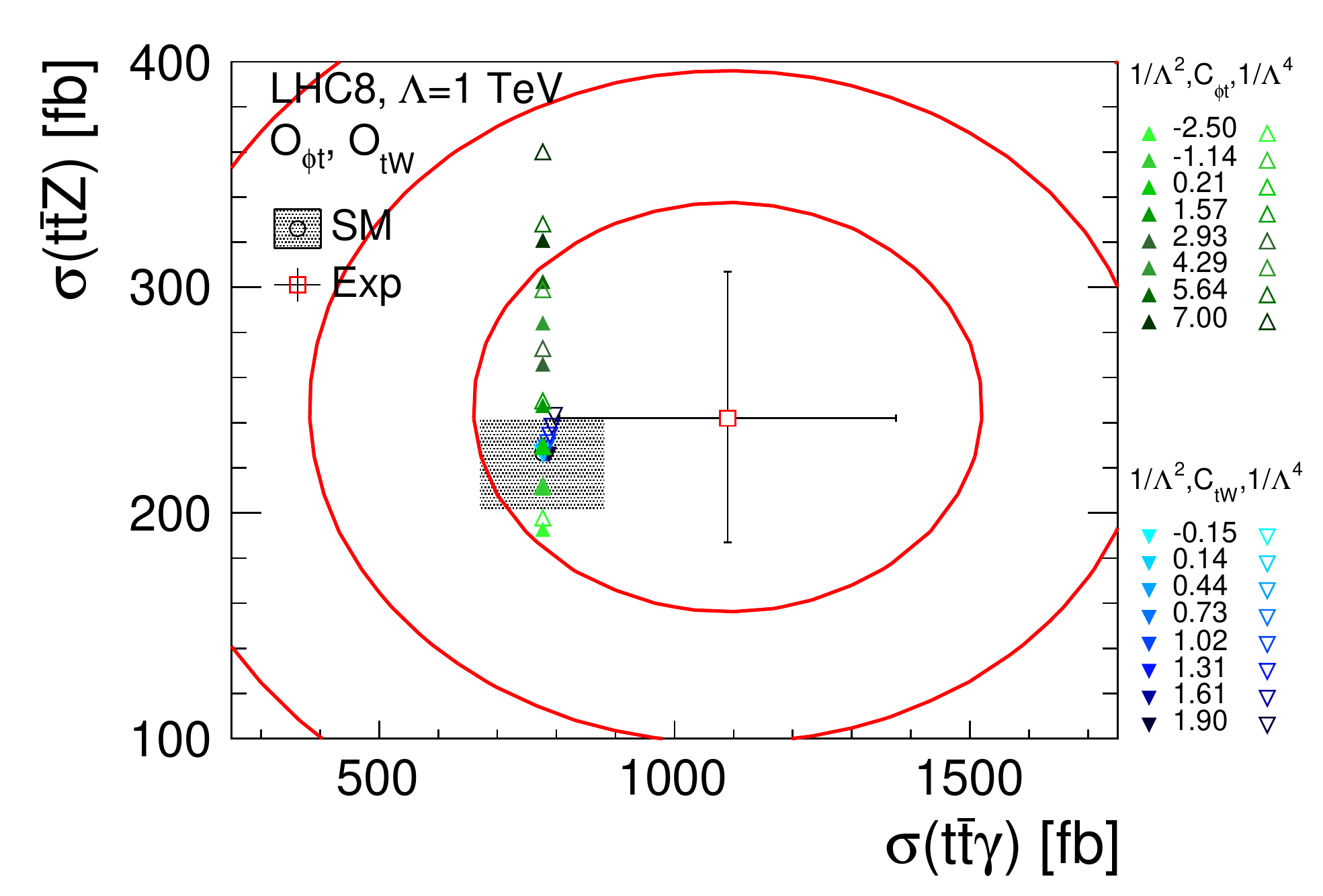}
\caption{Sensitivity of the $t\bar{t}\gamma$ and $t\bar{t}Z$ processes to the $\mathcal{O}_{\phi t}$ and $\mathcal{O}_{tW}$ operators. Details as in Fig.~\ref{fig:fq1tB}.}
\label{fig:fttw}
\end{figure}

Using the experimental measurements, one can further explore the sensitivity of
the $t\bar{t}\gamma$ and $t\bar{t}Z$ processes on the various operators as
shown in Figs.~\ref{fig:fq1tB}, \ref{fig:tgfq3} and \ref{fig:fttw}. In the
contour plots we include the experimental results of \cite{CMS:2014wma} for
$t\bar{t}\gamma$ and  \cite{Khachatryan:2015sha} for $t\bar{t}Z$ and the
corresponding one and two sigma contour plots. In this case, we assume there is
no correlation between the two measurements.
\noindent
The SM NLO predictions and the
corresponding scale uncertainties are also shown in the plots. We plot the
cross section obtained by varying the Wilson coefficients of the various
operators. For clarity and to avoid overcrowding the contour plots, we present
the operators in pairs. For the coefficients we employ the current constraints
to define our interval. Vertical lines indicate that the $t\bar{t}\gamma$ process
is not affected by the specific operator, i.e.~$\mathcal{O}_{\phi t},
\mathcal{O}^{(3)}_{\phi Q}$ and $\mathcal{O}^{(1)}_{\phi Q}$. Cross sections
with and without adding the
$\mathcal{O}(1/\Lambda^4)$ contributions from the squared EFT amplitudes are
compared. The $\mathcal{O}_{tB}$ operator is very loosely  constrained, and
therefore including the squared term for the large allowed values of the Wilson
coefficient has an enormous effect on the cross sections, as the
$\mathcal{O}(1/\Lambda^4)$ contribution scales like $C_{tB}^2$.  For the more
constrained current operators $\mathcal{O}^{(1)}_{\phi Q}$ and
$\mathcal{O}^{(3)}_{\phi Q}$, the squared contribution becomes important only
at the edges of the allowed intervals. 
\noindent
We also notice that for the
$\mathcal{O}_{\phi t}$ and $\mathcal{O}_{tG}$ operators the
$\mathcal{O}(1/\Lambda^4)$ contribution is important for a sizeable part of
allowed interval, in the first case because the constraints are rather loose
and in the second case because $\sigma^{(2)}_{tG}$ is large. Finally we note
that the contour plots qualitatively demonstrate the size of the experimental
uncertainties needed for these processes to have an impact on the allowed
values of the coefficients. 
\noindent
In that respect we observe for example that the
$\mathcal{O}_{tW}$ operator receives very stringent constraints from top decay,
and it is not expected to be further constrained by $t\bar{t}V$ measurements
even with a significant reduction of the experimental uncertainties. 

\section{Summary and conclusions}
\label{sec:conc}

We have presented the NLO QCD predictions in the SMEFT framework for the
associated production of a top-quark pair and a neutral gauge boson at the LHC.
In addition, we have considered top-pair production in $e^+e^-$ colliders and
the top loop-induced process $gg\to HZ$ at the LHC.  These processes are important
because they directly probe the neutral gauge-boson couplings to the top
quark, which are not well probed by other means.  In our approach we have
included the full set of dimension-six operators that parameterise these
couplings.

We have studied the contribution of each relevant dimension-six operator,
in both total cross sections and differential distributions.  We have presented
full results for $\mathcal{O}(\Lambda^{-2})$ contributions, along with
the squared contribution of each operator at $\mathcal{O}(\Lambda^{-4})$.
The latter contribution can be used to estimate uncertainties coming from higher order $\mathcal{O}(\Lambda^{-4})$ contributions.
Scale uncertainties are provided in all cases, and their reduction at NLO reflects the increased precision of our  predictions.

In $t\bar t\gamma$ and $t\bar tZ$, we find that the operator that contributes the most,
given our choice of operator normalisation, is the chromomagnetic one.  This
observation is particularly important in the context of a global EFT fit,
because it means that, when extracting information on operators modifying the top couplings with the
weak gauge bosons, uncertainties due to a possible non-vanishing chromomagnetic
operator should be carefully accounted for.  We also find that, the weak dipole
operators give extremely suppressed contributions at
$\mathcal{O}(1/\Lambda^2)$, due to a momentum suppression from the operator
structure, and in $t\bar tZ$ an additional accidental cancellation between $gg$
and $q\bar q$ initial states.  

A subset of the operators affects the associated production of the Higgs
with a $Z$ in gluon fusion, and we have considered their effects on this process
at the LHC.  This might provide additional constraints on the operators once
$ZH$ production is measured accurately at the LHC. Again, we find that the
contribution of $\mathcal{O}_{tG}$ is large, while 
all the current operators give the same contribution as they affect the axial
vector of the Z in the same way. The weak dipole operators do not contribute
due to charge conjugation parity.  We have also found that, at the ILC, $t\bar t$
production is sensitive to weak dipole operators, and could provide information
complementary to the LHC.

We have studied the sensitivity of the processes to the various operators 
in light of the current experimental measurements, as well as the constraints
currently placed on the operators from other top measurements and electroweak
precision observables. A discussion of the relevant uncertainties coming from missing higher orders in QCD and in the EFT has also been presented. The NLO results provide a solid basis for current and future measurements to be analysed in an EFT approach.

In summary, at NLO in QCD accuracy deviations from the SM in the top sector can be
extracted with improved accuracy and precision, keeping EFT uncertainties under
control.  As our calculation is based on the {\sc MG5\_aMC} framework,
matching with the parton shower and top decays with spin correlations can
be achieved in an automatic way.  Therefore, the corresponding simulations  can
be directly used in experimental analyses in the future to provide reliable information on possible EFT signals.
Furthermore, dedicated investigations of the features of deviations from the SM in these
processes can be performed based on our results, with an expected improvement in
sensitivity.

\acknowledgments
We acknowledge illuminating discussions with Christophe
Grojean, Alex Pomarol, Francesco Riva on the SMEFT and its range of validity.
We would like to thank Raoul~R\"ontsch and Markus~Schulze for discussions and
helpful checks.  C.Z.~would like to thank Valentin~Hirschi and Hua-Sheng~Shao
for valuable discussions about gauge anomaly.  This work has been performed in
the framework of the ERC Grant No.  291377 ``LHCTheory'' and has been supported
in part by the European Union as part of the FP7 Marie Curie Initial Training
Network MCnetITN (PITN-GA-2012-315877).  C.Z.~is supported by the United States
Department of Energy under Grant Contracts DE-SC0012704.

\appendix
\section{Connection with ``anomalous coupling'' approach}
In order to compare with
other work in the literature, we present here the connection of the Wilson
coefficients with the top quark anomalous couplings.

The anomalous coupling approach is followed in
\cite{Rontsch:2014cca,Rontsch:2015una} where the $t\bar{t}Z$ process is used to
probe anomalous top couplings.
Compared with the anomalous coupling parametrisation of the $\bar ttZ$ vertex,
\begin{equation}
	\lag_{ttZ}=e\bar{u}(p_t)\left[ 
		\gamma^\mu\left( C_{1,V}^Z+\gamma_5C_{1,A}^Z \right)
		+\frac{i\sigma^{\mu\nu}q_\nu}{m_Z}\left( C_{2,V}^Z+i\gamma_5C_{2,A}^Z \right)
	\right]v(p_{\bar t}) Z_{\mu}
	\label{ttZvertex}
\end{equation}
the relation between anomalous couplings and Wilson coefficients are:
\begin{align}
	C_{1,V}^Z&=\frac{1}{2}\left(C_{\varphi Q}^{(3)}-C_{\varphi Q}^{(1)}-C_{\varphi t}\right)
	\frac{m_t^2}{\Lambda^2 s_Wc_W}
	\\
	C_{1,A}^Z&=\frac{1}{2}\left(-C_{\varphi Q}^{(3)}+C_{\varphi Q}^{(1)}-C_{\varphi t}\right)
	\frac{m_t^2}{\Lambda^2 s_Wc_W}
	\label{eq:axialZcoup}
	\\
	C_{2,V}^Z&=\left( C_{tW} c_W^2-C_{tB} s_W^2 \right)\frac{2 m_t m_Z}{\Lambda^2s_Wc_W}
	\\
	C_{2,A}^Z&=0
\end{align}
	\\
Similar relations for the top photon interactions are:
\begin{equation}
	\lag_{tt\gamma}=e\bar{u}(p_t)\left[ 
		Q_t\gamma^\mu
		+\frac{i\sigma^{\mu\nu}q_\nu}{m_Z}\left( C_{2,V}^{\gamma}+i\gamma_5C_{2,A}^{\gamma} \right)
	\right]v(p_{\bar t}) A_{\mu}
\end{equation}

\begin{align}
	C_{2,V}^\gamma&=\left( C_{tW} +C_{tB} \right)\frac{2 m_t m_Z}{\Lambda^2}
	\\
	C_{2,A}^\gamma&=0
\end{align}
The CP-odd operators are zero simply because we have assumed $C_{tW}$ and $C_{tB}$
are real.

\section{Ratios for comparing with measurements}

\subsection{ATLAS - $t \bar t Z$} 
The ATLAS $t \bar t Z$ SM value, which is compared to the
measurement \cite{Aad:2015eua}, is calculated as follows
\begin{equation}
\sigma^{SM}_{ATLAS}(t \bar t Z) = \sigma^{SM}(t \bar t \ell^+ \ell^- , m(\ell \ell)>5 \text{ GeV}) + \sigma^{SM}(t \bar t Z) \times [1-BR(Z \rightarrow \ell^+ \ell^-)] \,.
\label{eq:RatiottZ}
\end{equation}
\noindent 
The $BR(Z \rightarrow \ell^+ \ell^-)$ is taken from {\sc MadSpin}
\cite{Artoisenet:2012st}. The branching ratio and the NLO cross sections including the absolute scale uncertainties, using our parameter settings, are

\begin{align*}
& \sigma^{SM}(t \bar t \mu^+ \mu^- , m(\ell \ell)>5 \text{ GeV}) = 11.63(1)^{ +1.00 }_{ -1.38 } \text{ fb} \\
& \sigma^{SM}(t \bar t \mu^+ \mu^- , m(\ell \ell)>10 \text{ GeV}) = 9.83(1)^{ +0.75 }_{ -1.13 } \text{ fb} \\
& \sigma^{SM}(t \bar t Z) = 226.5(6)^{ +15.1 }_{ -25.3 } \text{ fb} \\
& BR(Z \rightarrow \ell^+ \ell^-) = 0.1029\,.
\end{align*}
\noindent
Applying these results to Eq. \ref{eq:RatiottZ} we get 

\begin{equation*}
\sigma^{SM}_{ATLAS}(t \bar t Z) = 238.1(6)^{ +16.6 }_{ -26.8 } \text{ fb} \,.
\end{equation*}
\noindent
In order to compare our $t \bar t \mu^+ \mu^-$ results with the ATLAS
measurement we apply to the experimental result the $R^{t \bar t Z}_{ATLAS}$, defined as

\begin{equation*}
R^{t \bar t Z}_{ATLAS}=\frac{\sigma^{SM}(t \bar t \mu^+ \mu^- , m(\ell \ell)>10 \text{ GeV})}{\sigma^{SM}_{ATLAS}(t \bar t Z) } = 0.0413(1)^{ +0.0003 }_{ -0.0001 }\,.
\end{equation*}
\noindent
The corresponding value for 13 TeV is 
\begin{equation*}
R^{t \bar t Z, 13 \text{TeV}}_{ATLAS} = 0.0408(1)^{ +0.0003 }_{ -0.0002 }\,.
\end{equation*}

\subsection{CMS - $t \bar t \gamma$}
The measurement in \cite{CMS:2014wma} should be compared with the $W^+ b
W^- \bar b \gamma$ SM cross section calculated with $p_T(\gamma)> 20 \text{
GeV}$ and $\Delta R(\gamma, b/ \bar b)>0.1$. Our $t \bar t \gamma$ results are
with $p_T(\gamma)> 20 \text{ GeV}$, but they do not include photon radiation
from the $t ,\bar t $ decay products ($W^{\pm}, b, \bar b$). For this reason
the $R^{t \bar t \gamma}_{CMS}$ value is applied to the experimental result, defined at LO as follows
\begin{equation*}
R^{t \bar t \gamma}_{CMS}= \frac{\sigma^{SM}(t \bar t \gamma , p_T(\gamma)> 20 \text{ GeV})}{\sigma^{SM}(W^+ b W^- \bar b \gamma , p_T(\gamma)> 20 \text{ GeV} , \Delta R(\gamma, b/ \bar b)>0.1)} = 0.4531(4)^{ +0.0015 }_{ -0.0011 }\,.
\end{equation*}
The LO cross sections are
\begin{align*}
& \sigma^{SM}(t \bar t \gamma , p_T(\gamma)> 20 \text{ GeV}) = 604.0(3)^{ +234.1 }_{ -154.8 } \text{ fb} \\
& \sigma^{SM}(W^+ b W^- \bar b \gamma , p_T(\gamma)> 20 \text{ GeV} , \Delta R(\gamma, b/ \bar b)>0.1) = 1333.0(9)^{ +520.9 }_{ -344.9 } \text{ fb}\,.
\end{align*}
The corresponding value for 13 TeV is 

\begin{equation*}
R^{t \bar t \gamma, 13 \text{TeV}}_{CMS} = 0.4453(5)^{ +0.0008 }_{ -0.0003 }\,.
\end{equation*}

\bibliographystyle{JHEP}
\bibliography{ttZ_v4}

\end{document}